\documentclass{article}

\usepackage{PRIMEarxiv}

\RequirePackage{hyperref}

\usepackage{amsmath, graphicx, tabularx, multirow, enumerate, listings, subfigure, xcolor,comment}

\usepackage[utf8]{inputenc} 
\usepackage[T1]{fontenc}    
\usepackage{hyperref}       
\usepackage{url}            
\usepackage{booktabs}       
\usepackage{amsfonts}       
\usepackage{nicefrac}       
\usepackage{microtype}      
\usepackage{lipsum}
\usepackage{fancyhdr}       
\usepackage{graphicx}       
\graphicspath{{media/}}     

\title{Bayesian Methods in Tensor Analysis}

\author{
  Yiyao Shi \\
  University of California, Irvine \\
  Irvine, CA, US\\
  \texttt{yiyaos@uci.edu} \\
   \And
  Weining Shen \\
  University of California, Irvine \\
  Irvine, CA, US\\
  \texttt{weinings@uci.edu} \\
}

\begin{document}
\maketitle

\begin{abstract}
Tensors, also known as multidimensional arrays, are useful data structures in machine learning and statistics. In recent years, Bayesian methods have emerged as a popular direction for analyzing tensor-valued data since they provide a convenient way to introduce sparsity into the model and conduct uncertainty quantification. In this article, we provide an overview of frequentist and Bayesian methods for solving tensor completion and regression problems, with a focus on Bayesian methods. We review common Bayesian tensor approaches including model formulation, prior assignment, posterior computation, and theoretical properties. We also discuss potential future directions in this field. 
\end{abstract}

\keywords{Imaging analysis \and Posterior inference \and Recommender system \and Tensor completion \and Tensor decomposition \and Tensor regression}

\section{Introduction}

Tensors, also known as multidimensional arrays, are higher dimensional analogues of two-dimensional matrices. Tensor data analysis has gained popularity in many scientific research and business applications, including medical imaging~\cite{Bi2021}, recommender systems~\cite{Rendle2010}, relational learning~\cite{Trouillon2017}, computer vision~\cite{Song2019} and network analysis~\cite{Li2011}. There is a vast literature on studying tensor-related problems such as tensor decomposition~\cite{Kolda, Rabanser, Thanh}, tensor regression~\cite{Guhaniyogi2020R, Sun2014}, tensor completion~\cite{Song2019}, tensor clustering~\cite{Bi2021, Sun2014}, tensor reinforcement learning and deep learning~\cite{Sun2014}. Among them, tensor completion and tensor regression are two fundamental problems and we focus on their review in this article. 

Tensor completion aims at imputing missing or unobserved entries in a partially observed tensor. Important applications of tensor completion include providing personalized services and recommendations in context-aware recommender systems (CARS)~\cite{Rendle2010}, restoring incomplete images collected from magnetic resonance imaging (MRI) and computerized tomography (CT)~\cite{Gandy}, and inpainting missing pixels in images and videos~\cite{Liu2009, Mu}. In this review, we divide tensor completion methods into trace norm based methods and decomposition based methods, and introduce common approaches in each category.

Different from tensor completion, tensor regression investigates the association between tensor-valued objects and other variables. For example, medical imaging data such as brain MRI are naturally stored as a multi-dimensional array, and tensor regression methods are applied to analyze their relationship with clinical outcomes (e.g., diagnostic status, cognition and memory score)~\cite{Li2017, Sun2017}. Based on the role that the tensor-valued object plays in the regression model, tensor regression methods can be categorized into tensor predictor regression and tensor response regression.  


Frequentist approaches have been successful in tensor analysis \cite{Wimalawarne,Bi2021}. In recent years, Bayesian approaches have also gained popularity as they provide a useful way to induce sparsity in tensor models and conduct uncertainty quantification for estimation and predictions. In this article, we will briefly discuss common frequentist approaches to solve tensor completion and regression problems and focus on Bayesian approaches. We also review two commonly used tensor decompositions, i.e., CANDECOMP/PARAFAC (CP) decomposition~\cite{CP} and the Tucker decomposition~\cite{Tucker}, since they are the foundations for most Bayesian tensor models. For example, many Bayesian tensor completion  approaches begin with certain decomposition structure on the tensor-valued data and then use Bayesian methods to infer the decomposition parameters and impute the missing entries. Based on the decomposition structures being utilized, we divide these methods into CP-based, Tucker-based, and nonparametric methods. For tensor regression methods, we classify the Bayesian tensor regression into Bayesian tensor predictor regression and Bayesian tensor response regression. For each category, we review the prior construction, model setup, posterior convergence property and sampling strategies. 

\begin{figure*}[htp]
	{\centering
	\includegraphics[width=\linewidth]{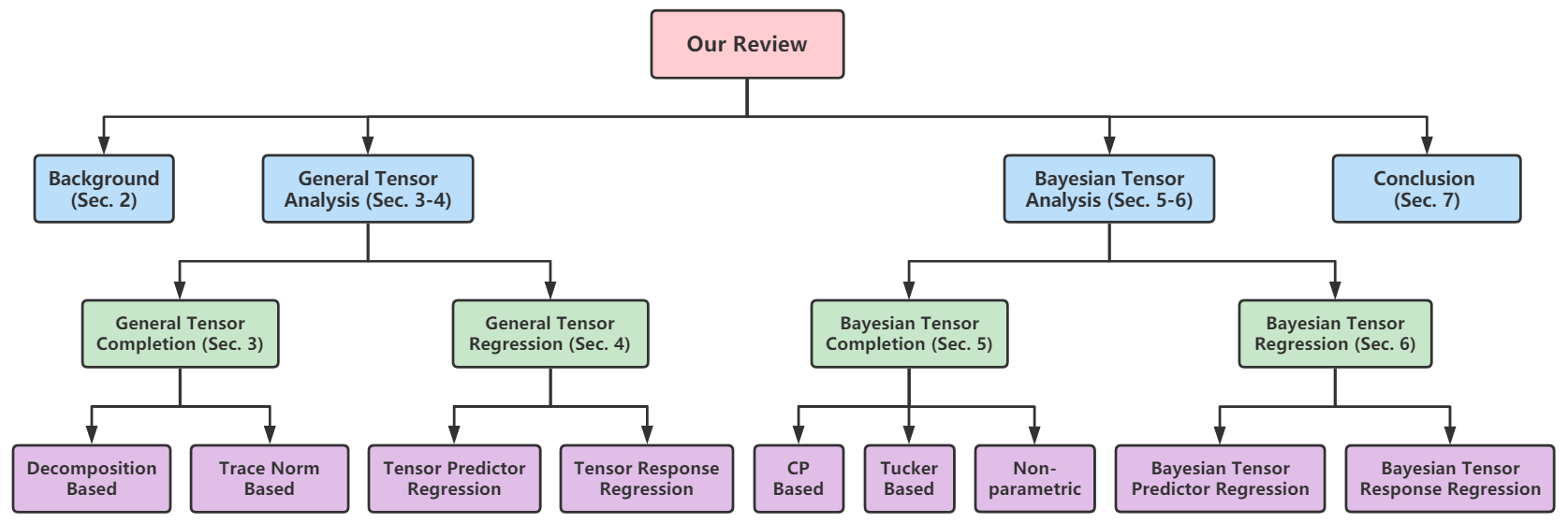}
		\caption{Outline of this survey.}
		\label{outline}}
\end{figure*}

The rest of this article is organized as follows. Section~\ref{BG} provides a background introduction to tensor notations, operations and decompositions. Section~\ref{TC} and  \ref{TR} review common frequentist approaches for tensor completion and regression problems, respectively. Section~\ref{BD} and \ref{BR} review  Bayesian tensor completion and regression approaches, including the prior construction, posterior computing, and theoretical properties. Section~\ref{Conclusion} provides concluding remarks and discusses several future directions for Bayesian tensor analysis. Figure~\ref{outline} shows an outline of our review.

\section{Background}\label{BG}
In this section, we follow~\cite{Kolda} and introduce notation, definitions, and operations related to tensors. We also discuss two popular tensor decomposition approaches and highlight some challenges in tensor analysis. 

\begin{figure}[htp]
	{\centering
	\includegraphics[width=.7\linewidth]{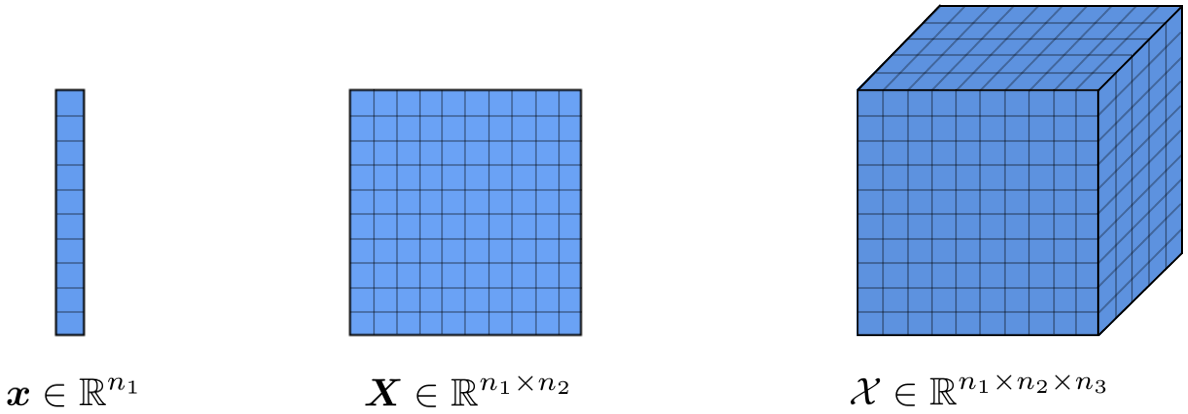}
		\caption{An example of first, second and third-order tensors.}
		\label{f1}}
\end{figure}

\subsection{Basics}
\begin{figure*}[htp]
	\centering
	\subfigure[Mode-1 (column) fibers: $\mathcal{X}_{:i_2i_3}$]{
		\begin{minipage}[t]{0.35\linewidth}
			\centering
			\includegraphics[width=\linewidth]{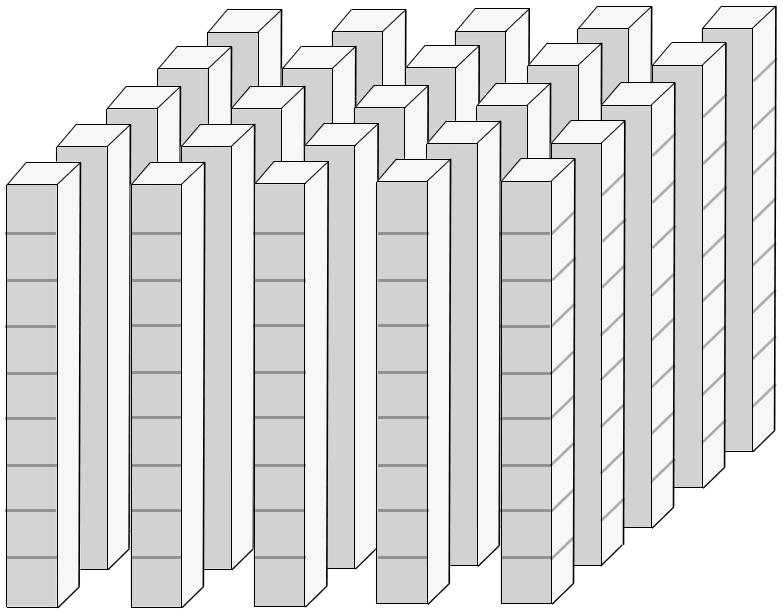}
	\end{minipage}}
	\subfigure[Mode-2 (row) fibers: $\mathcal{X}_{i_1:i_3}$]{
		\begin{minipage}[t]{0.29\linewidth}
			\centering
			\includegraphics[width=\linewidth]{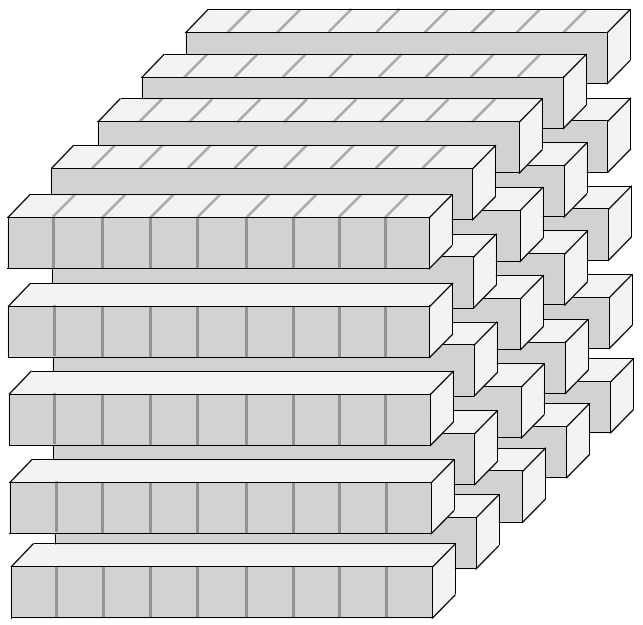}
	\end{minipage}}
	\subfigure[Mode-3 (tube) fibers: $\mathcal{X}_{i_1i_2:}$]{
		\begin{minipage}[t]{0.26\linewidth}
			\centering
			\includegraphics[width=\linewidth]{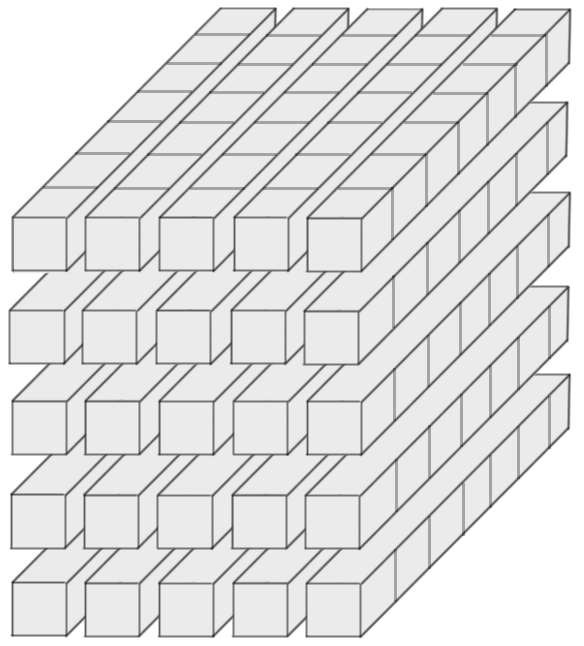}
	\end{minipage}}
	\subfigure[Horizontal slices: $\mathcal{X}_{i_1::}$]{
		\begin{minipage}[t]{0.35\linewidth}
			\centering
			\includegraphics[width=\linewidth]{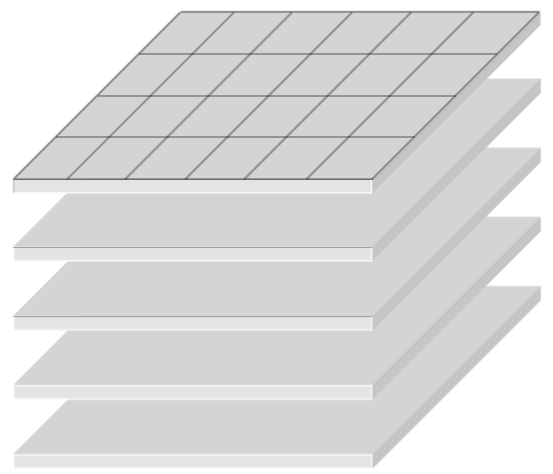}
	\end{minipage}}
	\subfigure[Lateral slices: $\mathcal{X}_{:i_2:}$]{
		\begin{minipage}[t]{0.29\linewidth}
			\centering
			\includegraphics[width=\linewidth]{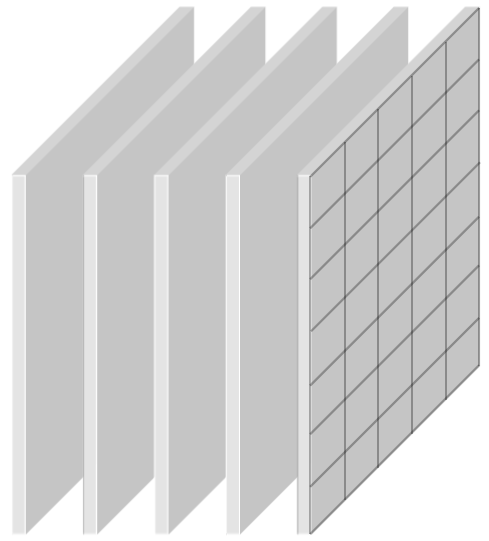}
	\end{minipage}}
	\subfigure[Frontal slices: $\mathcal{X}_{::i_3}$]{
		\begin{minipage}[t]{0.3\linewidth}
			\centering
			\includegraphics[width=\linewidth]{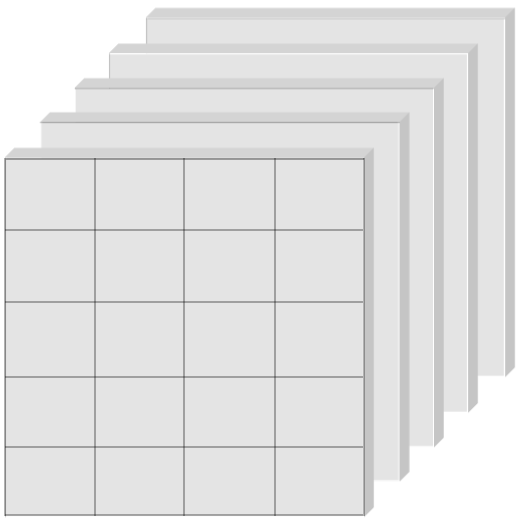}
	\end{minipage}}
	\caption{Example of fibers and slices of third-order tensor. This figure is reproduced based on Figure 2.1 and 2.2 in~\cite{Kolda}.}
	\label{f2}
\end{figure*}

\paragraph*{Notation:} A tensor is a multidimensional array. The dimension of a tensor is also known as {\it mode}, {\it way}, or {\it order}. A first-order tensor is a vector; a second-order tensor is a matrix; and tensors of order three and higher are referred to as higher-order tensors (see Figure~\ref{f1}). In this review, a tensor is denoted by Euler script letter $\mathcal{X}\in \mathbb{R}^{n_1\times n_2\times...\times n_d}$. Here $d$ is the order of tensor $\mathcal{X}$, and $n_k$ is the marginal dimension of the $k$th mode ($k=1,2,...,d$). The $(i_1,i_2,...,i_d)$th element of the tensor $\mathcal{X}$ is denoted by $x_{i_1i_2...i_d}$ for $i_k=1,2,...,n_k$ and $k=1,2,...,d$. Subarrays of a tensor are formed through fixing a subset of indices in the tensor. A {\it fiber} is a vector defined by fixing all but one indices of a tensor, and a {\it slice} is a matrix created by fixing all the indices except for those of two specific orders in the tensor. For instance, a third-order tensor $\mathcal{X}\in \mathbb{R}^{n_1\times n_2\times n_3}$ has column, row and tube fibers, which are respectively denoted by $\mathcal{X}_{:i_2i_3}, \mathcal{X}_{i_1:i_3}$, and $\mathcal{X}_{i_1i_2:}$ (see Figure~\ref{f2}(a)(b)(c)). A third-order tensor also has horizontal, lateral, and frontal slices, denoted by $\mathcal{X}_{i_1::}, \mathcal{X}_{:i_2:}$ and $\mathcal{X}_{::i_3}$, respectively (see Figure~\ref{f2}(d)(e)(f)).

\begin{figure*}[htp]
	{\centering
	\includegraphics[width=.9\linewidth]{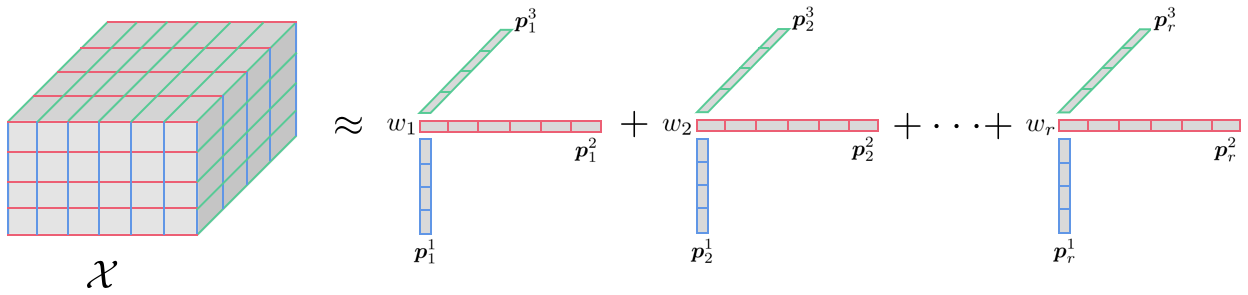}
		\caption{Rank-$r$ CP decomposition for a third-order tensor: $\mathcal{X}\approx \sum_{j=1}^r w_j\boldsymbol{p}_j^1\circ\boldsymbol{p}_j^2\circ\boldsymbol{p}_j^3.$}
		\label{f3}}
\end{figure*}

\paragraph*{Tensor Operations:} Here we introduce some tensor operations following~\cite{Kolda}. The {\it norm} of a tensor $\mathcal{X}\in \mathbb{R}^{n_1\times n_2\times...\times n_d}$ is defined as the square root of the sum of the squares of all elements, i.e.,
\begin{equation}
    \Vert\mathcal{X}\Vert = \sqrt{\sum_{i_1=1}^{n_1}\sum_{i_2=1}^{n_2}\cdots\sum_{i_d=1}^{n_d}x_{i_1i_2...i_d}^2}.
    \label{norm}
\end{equation}
For two same-sized tensors $\mathcal{X}, \mathcal{Y}\in\mathbb{R}^{n_1\times...\times n_d}$, their {\it inner product} is the sum of products of their corresponding entries, i.e.,
\begin{equation}
    \langle\mathcal{X}, \mathcal{Y}\rangle =\sum_{i_1=1}^{n_1}\sum_{i_2=1}^{n_2}\cdots\sum_{i_d=1}^{n_d} x_{i_1i_2...i_d}y_{i_1i_2...i_d}.
    \label{inner product}
\end{equation}
It immediately follows that $\langle\mathcal{X},\mathcal{X}\rangle=\Vert\mathcal{X}\Vert^2.$ The {\it tensor Hadamard product} of two tensors $\mathcal{X}\in\mathbb{R}^{n_1\times...\times n_d}$ and $\mathcal{Y}\in\mathbb{R}^{n_1\times...\times n_d}$ is denoted by $\mathcal{X}*_H\mathcal{Y}\in\mathbb{R}^{n_1\times...\times n_d}$; each entry of $\mathcal{X}*_H\mathcal{Y}$ is the product of the corresponding entries in tensors $\mathcal{X}$ and $\mathcal{Y}$:
\begin{equation}
    (\mathcal{X}*_H\mathcal{Y})_{i_1...i_d}=x_{i_1...i_d}\cdot y_{i_1...i_d}.
    \label{Hadamard}
\end{equation}
The {\it tensor contraction product}, also known as the {\it Einstein product}, of two tensors $\mathcal{X}\in\mathbb{R}^{n_1\times...\times n_d\times p_1\times...\times p_k}$ and $\mathcal{Y}\in\mathbb{R}^{p_1\times...\times p_k\times m_1\times...\times m_q}$ is denoted by $\mathcal{X}*\mathcal{Y}\in\mathbb{R}^{n_1\times...\times n_d\times m_1\times...\times m_q}$ and defined as
\begin{equation}
(\mathcal{X}*\mathcal{Y})_{i_1,...,i_d,j_1,...,j_q}=\sum_{c_1=1}^{p_1}\cdots\sum_{c_k=1}^{p_k}x_{i_1,...,i_d,c_1,...,c_k}y_{c_1,...,c_k,j_1,...,j_q},
\label{contraction product}
\end{equation}
where $i_g=1,2,...,n_g$ for $g=1,2,...,d$, and $j_s=1,2,...,m_s$ for $s=1,2,...,q$.
Moreover, a $d$th-order tensor $\mathcal{X}\in \mathbb{R}^{n_1\times n_2\times...\times n_d}$ is {\it rank one} if it can be written as the outer product of $d$ vectors, i.e,
$$\mathcal{X}=\boldsymbol{p}^1\circ \boldsymbol{p}^2\circ\cdots\circ \boldsymbol{p}^d,$$
where $\boldsymbol{p}^k=(p_1^k,p_2^k,...,p_{n_k}^k)\in\mathbb{R}^{n_k}$ $(k=1,2,...,d)$ is a vector, and the symbol ``$\circ$'' represents the vector outer product. It means that each element of the tensor $\mathcal{X}$ is the product of corresponding vector elements: $x_{i_1i_2...i_d}=p_{i_1}^1p_{i_2}^2...p_{i_d}^d$ for $i_k=1,2,...,n_k$ and $k=1,2,...,d$. A tensor $\mathcal{X}$ is {\it rank $r$} if $r$ is the smallest number such that $\mathcal{X}$ is the sum of $r$ outer products of vectors: $\mathcal{X}=\sum_{j=1}^r\boldsymbol{p}_j^1\circ\boldsymbol{p}_j^2\circ\cdots\circ\boldsymbol{p}_j^d$.

Tensor {\it matricization}, also known as tensor {\it unfolding} or {\it flattening}, is an operation that transforms a tensor into a matrix. Given a tensor $\mathcal{X}\in\mathbb{R}^{n_1\times n_2\times...\times n_d}$, the $k$th-mode matricization arranges the mode-$k$ fibers to be columns of the resulting matrix, which is denoted by $\boldsymbol{X}_{(k)}$ ($k=1,2,...,d$). The element $(i_1,i_2,...,i_d)$ of tensor $\mathcal{X}$ corresponds to the entry $(i_k,j)$ of $\boldsymbol{X}_{(k)}$, where $j=1+\sum_{t=1,t\neq k}^d (i_t-1)J_t$ with $J_t=\prod_{m=1,m\neq k}^{t-1}n_m$. In addition, a tensor can be transformed into a vector through tensor {\it vectorization}. For a tensor $\mathcal{X}\in\mathbb{R}^{n_1\times...\times n_d}$, the vectorization of $\mathcal{X}$ is denoted by vec($\mathcal{X})\in\mathbb{R}^{\prod_{i=1}^d n_i}$. The element $(i_1,i_2,...,i_d)$ of tensor $\mathcal{X}$ corresponds to the element $1+\sum_{t=1}^d (i_t-1)M_t$ of vec($\mathcal{X}$), where $M_t=\prod_{m=1}^{t-1}n_m$.

The {\it $k$-mode tensor matrix product} of a tensor $\mathcal{X}\in\mathbb{R}^{n_1\times n_2\times \cdots \times n_d}$ with a matrix $\boldsymbol{A}\in\mathbb{R}^{m \times n_k}$ is denoted by $\mathcal{X}\times_k\boldsymbol{A}$, which is of size $n_1\times \cdots  \times n_{k-1}\times m\times n_{k+1}\times \cdots 
 \times n_d$. Elementwise, we have $(\mathcal{X}\times_k\boldsymbol{A})_{i_1,\ldots,i_{k-1},j,i_{k+1},\ldots,i_d}=\sum_{i_k=1}^{n_k}\mathcal{X}_{i_1,\ldots,i_d}\boldsymbol{A}_{ji_k}$. The {\it $k$-mode vector product} of a tensor $\mathcal{X}\in\mathbb{R}^{n_1\times n_2\times \cdots \times n_d}$ with a vector $\boldsymbol{a}\in\mathbb{R}^{n_k}$ is denoted by $\mathcal{X}\bar{\times}_k\boldsymbol{a}$, which is of size $n_1\times \cdots \times n_{k-1}\times n_{k+1}\times\cdots \times n_d$. Elementwise, $(\mathcal{X}\bar{\times}_k\boldsymbol{a})_{i_1\ldots i_{k-1}i_{k+1}\ldots i_d}=\sum_{i_k=1}^{n_k}x_{i_1i_2...i_d}a_{i_k}.$

\subsection{Tensor Decompositions}\label{Decomposition}
Tensor decompositions refer to methods that express a tensor by a combination of simple arrays. Here we introduce two widely-used tensor decompositions and discuss their applications.  

\paragraph*{CP decomposition:} The {\it CANDECOMP/PARAFAC decomposition} ({\it CP decomposition})~\cite{CP} factorizes a tensor into a sum of rank-1 tensors. For a $d$th-mode tensor $\mathcal{X}$, the rank-$r$ CP decomposition is written as
\begin{equation}
    \mathcal{X}\approx \sum_{j=1}^r w_j\boldsymbol{p}_j^1\circ\boldsymbol{p}_j^2\circ\cdots\circ\boldsymbol{p}_j^d,
    \label{CP}
\end{equation}
where $w_j\in\mathbb{R}, \boldsymbol{p}_j^k\in\mathbb{S}^{n_k-1}, j=1,...,r, k=1,2,...,d, \mathbb{S}^{n_k-1}=\{\boldsymbol{a}\in\mathbb{R}^{n_k}|\Vert\boldsymbol{a}\Vert=1\},$ and $\circ$ is the outer product. See Figure~\ref{f3} for a graphical illustration of CP decomposition. Sometimes the CP-decomposition is denoted by an abbreviation: $\mathcal{X}\approx [\![\boldsymbol{W}; \boldsymbol{P}^1,\boldsymbol{P}^2,...,\boldsymbol{P}^d]\!],$ where $\boldsymbol{W}=\text{diag(}w_1,...,w_r)\in\mathbb{R}^{r\times r}$ is a diagonal matrix, and $\boldsymbol{P}^k=[\boldsymbol{p}_1^k,\boldsymbol{p}_2^k...,\boldsymbol{p}_r^k]\in\mathbb{R}^{n_k\times r}$ are factor matrices. If tensor $\mathcal{X}$ admits a CP structure, then the number of free parameters changes from $\prod_{i=1}^d n_i$ to $r\times(\sum_{i=1}^d n_i-d+1)$.

If Equation (\ref{CP}) attains equality, the decomposition is called an exact CP decomposition. Even for an exact CP decomposition, there is no straightforward algorithm to determine the rank $r$ of a specific tensor, and in fact the problem is NP-hard~\cite{Hastad}. In practice, most procedures numerically infer the rank by fitting CP models with different ranks and choosing the one with the best numerical performance. 

\begin{figure}[htp]
	{\centering
	\includegraphics[width=.6\linewidth]{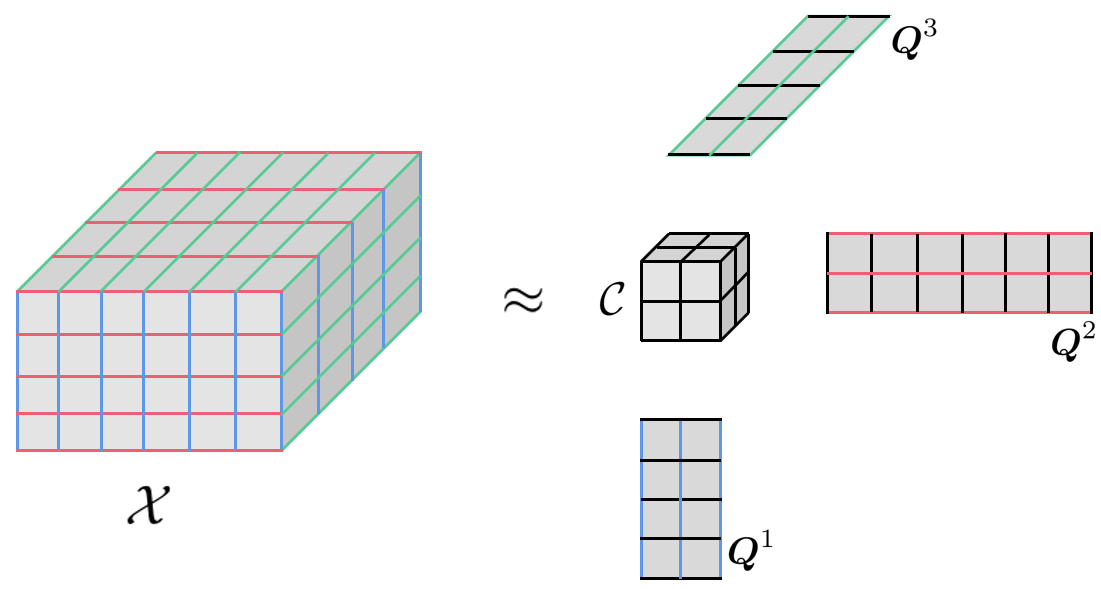}
		\caption{Tucker decomposition of the third-order tensor $\mathcal{X}\in\mathbb{R}^{n_1\times n_2\times n_3}$, where $\mathcal{C}\in\mathbb{R}^{m_1\times m_2\times m_3}$ is the core tensor, and $\boldsymbol{Q}^k\in \mathbb{R}^{n_k\times m_k} (k=1,2,3)$ are factor matrices.}
		\label{f4}}
\end{figure}

\paragraph*{Tucker decomposition:}
The {\it Tucker decomposition} factorizes a tensor into a core tensor multiplied by a matrix along each mode. Given a $d$th-order tensor $\mathcal{X}\in\mathbb{R}^{n_1\times n_2\times...\times n_d}$, the Tucker decomposition is defined as
\begin{equation}
    \mathcal{X}\approx \mathcal{C}\times_1 \boldsymbol{Q}^1\times_2\boldsymbol{Q}^2\times_3\cdots\times_d\boldsymbol{Q}^d=\sum_{j_1=1}^{m_1}\sum_{j_2=1}^{m_2}\cdots\sum_{j_d=1}^{m_d}c_{j_1j_2\dots j_d}\boldsymbol{q}_{j_1}^1\circ\boldsymbol{q}_{j_2}^2\circ\cdots\circ\boldsymbol{q}_{j_d}^d,\\
\label{Tucker}\end{equation}
where $\mathcal{C}\in\mathbb{R}^{m_1\times m_2\times...\times m_d}$ is the core tensor, $\boldsymbol{Q}^k\in\mathbb{R}^{n_k\times m_k}(k=1,2,...,d)$ are factor matrices, $c_{j_1j_2...j_d}\in\mathbb{R}, \boldsymbol{q}_{j_k}^k\in\mathbb{S}^{n_k-1} (j_k=1,2,...,m_k,k=1,2,...,d)$. See Figure~\ref{f4} for a graphical illustration of Tucker decomposition. The Tucker decomposition can be denoted as $\mathcal{X}\approx [\![\mathcal{C}; \boldsymbol{Q}^1, \boldsymbol{Q}^2,...,\boldsymbol{Q}^d]\!].$ If $\mathcal{X}$ admits a Tucker structure, the number of free parameters in $\mathcal{X}$ changes from $\prod_{i=1}^d n_i$ to $\sum_{i=1}^d (n_i-1)\times m_i+\prod_{i=1}^d m_i$.

The {\it $k$-rank} of $\mathcal{X}\in\mathbb{R}^{n_1\times...\times n_d}$, denoted by rank$_k$($\mathcal{X}$), is defined as the column rank of $k$th-mode matricization matrix $\boldsymbol{X}_{(k)}$. Let $R_k=$rank$_k(\mathcal{X})$, then $\mathcal{X}$ is a rank-$(R_1,R_2,...,R_d)$ tensor. Trivially, $R_k\leq n_k$ for $k=1,2,...,d$. When the equality in 
 Equation (\ref{Tucker}) is attained, the decomposition is called an exact Tucker decomposition. For a given tensor $\mathcal{X}$, there always exists an exact Tucker decomposition with core tensor $\mathcal{C}\in\mathbb{R}^{m_1\times m_2\times\cdots\times m_d}$ where $m_k$ is the true $k$-rank for $k=1,2,...,d$. Nevertheless, for one or more $k$, if $m_k<R_k$, then the Tucker decomposition is not necessarily exact; and if $m_k>R_k$, the model will contain redundant parameters. Therefore, we usually want to identify the true tensor rank, i.e., $m_k=R_k$. While this job is easy for noiseless complete tensors, for tensors obtained in real-world applications, which are usually noisy or partially observed, the rank still needs to be determined by certain searching procedures. 

\subsection{Challenges in tensor analysis}
In tensor analysis, the ultrahigh dimensionality of the tensor-valued coefficients and tensor data creates challenges such as heavy computational burden and vulnerability to model overfitting. Conventional approaches usually transform the tensors into vectors or matrices and utilize dimension reduction and low-dimensional techniques. However, these methods are usually incapable of accounting for the dependence structure in tensor entries. In the past decades, an increasing number of studies have imposed decomposition structures on the tensor-valued coefficients or data; thus naturally reducing the number of free parameters, and avoiding the issues brought by high dimensionality.

In this paper, we focus on tensor regression and tensor completion problems, where various decomposition structures including CP and Tucker have been widely used. Specifically, a large proportion of tensor completion methods are realized through inferring the decomposition structure based on the partially observed tensor, and then impute the missing values through the inferred decomposition structure. Also, tensor regression problems usually include tensor-valued coefficients, and decomposition structures are imposed on the coefficient tensor to achieve parsimony in parameters. In both situations, the decomposition is not performed on a completely observed tensor, thus the rank of the decomposition cannot be directly inferred from the data. Most optimization-based approaches determine the rank by various selection criteria, which may suffer from low stability issues. Bayesian approaches perform automatic rank inference through the introduction of sparsity-inducing priors. However, efficient posterior computing and study of theoretical properties of the posterior distributions are largely needed. 

Low rankness and sparsity are commonly used assumptions in the literature to help reduce the number of free parameters. For non-Bayesian methods, oftentimes the task is formulated into an optimization problem, and the assumptions are enforced by sparsity-inducing penalty functions. In comparison, the Bayesian methods perform decompositions in the probabilistic setting, and enforce sparsity assumptions through sparsity priors. We will discuss more details about these approaches and how they resolve challenges in the following sections.

\section{Tensor Completion}\label{TC}
Tensor completion methods aim at imputing missing or unobserved entries from a partially observed tensor. It is a fundamental problem in tensor research  and has wide applications in numerous domains. For instance, tensor completion techniques are extensively utilized in context-aware recommender systems (CARS) to provide personalized services and recommendations~\cite{Karatzoglou, Bi2018, Tarzanagh2022}. In ordinary recommender systems, 
the user-item interaction data are collected and formulated into a sparse interaction matrix, and the goal is to complete the matrix and thus recommend individualized items to the users. In CARS, the user-item interaction is collected with their contextual information (e.g., time and network), and the data are formulated as a high-order tensor where the modes respectively represent users, items, and contexts~\cite{Adomavicius2010}. Therefore, the matrix completion problem in ordinary recommender systems is transformed into a tensor completion problem in CARS, and the purpose is to make personalized recommendations to users based on the collected user-item interaction and contextual information.

\begin{figure*}[htp]
	{\centering
	\includegraphics[width=\linewidth]{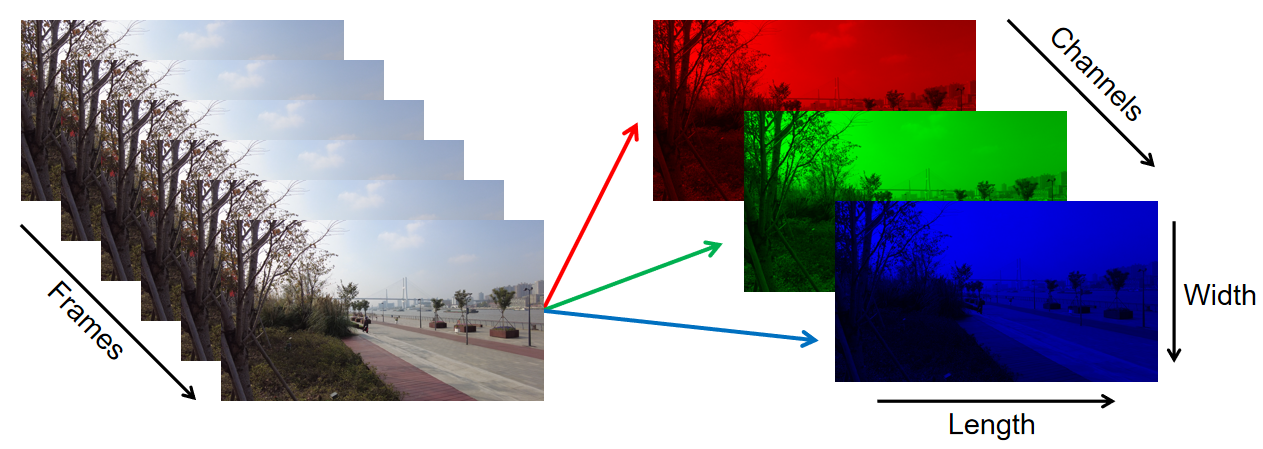}
		\caption{An illustration of color videos. Each frame of the video is formulated as a third-order tensor, where the modes are length, width and channels (RGB channels in this case). The frames are then stacked into a fourth-order tensor according to time order.}
	\label{f7}}
\end{figure*}

Apart from CARS, tensor completion is also applied in other research domains including healthcare, computer vision and chemometrics~\cite{Song2019}. For example, medical images collected from MRI and CT play important roles in the clinical diagnosis process. Due to the high acquisition speed, oftentimes these high-order images are incomplete, thus necessitating the application of tensor completion algorithms~\cite{Gandy, Bazerque2013}. In the field of computer vision, color videos can be represented by a fourth-order tensor (length$\times$width$\times$channel$\times$frame) by stacking the frames in time order (see Figure~\ref{f7}). Tensor completion can be adopted to impute the missing pixels and restore the lossy videos~\cite{Liu2009, Mu}. As another example, chemometrics is a discipline that employs mathematical, statistical and other methods to improve chemical analysis. Tensor completion methods have been successfully applied on various benchmark chemometric datasets including semi-realistic amino acid fluorescence datasets~\cite{Bro1997} and flow injection datasets~\cite{Narita}.

Tensor completion can be viewed as a generalization of matrix completion. Since the matrix completion problems have been well-studied in the past few decades, a natural way to conduct tensor completion is to unfold or slice the tensor into a matrix (or matrices) and apply matrix completion methods to the transformed matrix (or matrices). Nevertheless, the performance and efficiency of such approaches are largely reduced by the loss of structural information during the matricization process and excessive computational cost due to the high dimensionality of the original tensor. 

Under such circumstances, various methods that specifically focus on high-order tensor completion have been developed. Among these techniques, a classical group of approaches perform tensor completion through tensor decomposition. Generally speaking, these methods impose a decomposition structure on a tensor, and estimate the decomposition parameters based on the observed entries of the tensor. After that, the estimated decomposition structure is utilized to infer the missing entries of the tensor. Trace-norm based methods are another popular class of tensor completion methods. These methods first formulate tensor completion as a rank minimization problem, and then employ the tensor trace norm to further transform the task into a convex optimization problem. Finally, various optimization techniques are applied to solve the problem and thus complete the tensor. In this section we provide a brief review of decomposition based and trace norm based tensor completion methods. More details on these two methods and other variants of tensor completion approaches can be found in Song et al.~\cite{Song2019}. 

\subsection{Decomposition Based Methods} \label{DE}
CP decomposition \eqref{CP} and Tucker decomposition \eqref{Tucker} are two of the most commonly used decomposition-based methods for  tensor completion. In \cite{Tomasi}, the authors propose to perform CP decomposition on partially observed tensors by iteratively imputing the missing values and estimating the latent vectors in the CP structure. Specifically, in iteration $s~(s\geq 1)$, the partially observed tensor $\mathcal{X}$ is  completed by:
$$ \tilde{\mathcal{X}}^{(s)}=\mathcal{X}*_{H}\mathcal{M}+\mathcal{Y}^{(s)}*_H(\boldsymbol{1}-\mathcal{M}),$$
where $*_H$ is the tensor Hadamard product defined in \eqref{Hadamard}, $\tilde{\mathcal{X}}^{(s)}, \mathcal{X}, \mathcal{Y}^{(s)}, \mathcal{M}\in\mathbb{R}^{n_1\times...\times n_d}$ are tensors of same size, $\tilde{\mathcal{X}}^{(s)}$ is the completed tensor, $\mathcal{Y}^{(s)}$ is the interim low-rank approximation based on CP decomposition, and $\mathcal{M}$ is the observation index tensor defined as
$$ \mathcal{M}_{i_1...i_d}=\begin{cases} 1\quad\text{if }\mathcal{X}_{i_1...i_d}\text{ is observed},\\ 0\quad\text{if }\mathcal{X}_{i_1...i_d}\text{ is unobserved}.\end{cases}$$

After the tensor is completed, the decomposition parameters are estimated by alternating least square optimization (ALS). The loop of tensor completion and parameter estimation is repeated until convergence. 

Similar approaches were adopted by Kiers et al.~\cite{Kiers} and Kroonenberg~\cite{Kroonenberg} to impute missing entries. These methods are referred to as EM-like methods, because they can be viewed as a special expectation maximization (EM) method when the residuals independently follow a Gaussian distribution. While the EM-like methods are usually easy to implement, they may not perform well (e.g., slow convergence and converging to a local maximum)  when there is a  high proportion of missing values. 

Also based on the CP decomposition, Bro et al.~\cite{Bro1998} propose another type of tensor completion method called the Missing-Skipping (MS) method. It conducts the CP decomposition based only on the observed entries in the tensor, and is typically more robust than the EM-like approaches when applied to tensors with a high proportion of missingness. In general, the MS methods seek to optimize the following objective function
\begin{equation}
    L=\sum_{(i_1,i_2,...,i_d)\in\Omega}\mathcal{D}(\mathcal{X}_{i_1,...i_d},\mathcal{Y}_{i_1,...,i_d}),
    \label{eq5}
\end{equation}
where $\mathcal{X}\in\mathbb{R}^{n_1\times...n_d}$ is the observed tensor, $\mathcal{Y}\in\mathbb{R}^{n_1\times...\times n_d}$ is the estimated tensor with a CP structure, $\Omega$ is a set containing indices of all observed entries in tensor $\mathcal{X}$, and $\mathcal{D}$ is an error measure. 

Under the optimization framework~(\ref{eq5}), Tomasi and Bro~\cite{Tomasi} define the error measure $\mathcal{D}$ to be the squared difference between the observed and estimated entry $\mathcal{D}(\mathcal{X}_{i_1,...i_d},\mathcal{Y}_{i_1,...,i_d})=(\mathcal{X}_{i_1,...i_d}-\mathcal{Y}_{i_1,...,i_d})^2$, and employ a modified Gauss-Newton iterative algorithm (i.e., Levenberg-Marquardt method)~\cite{Levenberg, Marquardt} to solve the optimization problem. Acar et al.~\cite{Acar} utilize a weighted error and minimize the objective function based on the first-order gradient, which is shown to be more scalable to larger problem sizes than the second-order optimization method in \cite{Tomasi}. Moreover, the optimization problem can be analyzed in a Bayesian setting by treating the error measure $\mathcal{D}$ to be the negative log-likelihood function. We will discuss more details about these probabilistic methods  in Section~\ref{BD}.

Tucker decomposition is another widely utilized tool to conduct tensor completion. While the CP-based completion approaches enjoy nice properties including uniqueness (with the exception of elementary indeterminacies of scaling and permutation) and nice interpretability of latent vectors, methods that employ Tucker structure are able to accommodate more complex interaction among latent vectors and are more effective than CP-based methods. Therefore, in some real-world applications where the completion accuracy is prioritized over the uniqueness and latent vector interpretation, Tucker-based approaches are potentially more suitable than the CP-based methods.

Similar to CP-based methods, EM-like approaches and MS approaches are still two conventional ways for Tucker-based tensor completion algorithms. Walczak and Massart~\cite{Walczak} and Andersson and Bro~\cite{Andersson} discuss the idea of utilizing EM-like Tucker decomposition to solve tensor completion in their earlier works. This method is further combined with higher-order orthogonal iteration to impute missing data~\cite{Geng}. As an example of MS Tucker decomposition, Karatzoglou et al.~\cite{Karatzoglou} employ a stochastic gradient descent algorithm to optimize the loss function based only on the observed entries. There are also researches that develop MS-based methods under a Bayesian framework. See Section~\ref{BD} for more details.

In recent years, several studies utilize hierarchical tensor (HT) representations to provide a generalization of classical Tucker models. Most of the HT representation based methods are implemented using projected gradient methods. For instance, Rauhut et al.~\cite{Rauhut2015, Rauhut2017} employ a Riemannian gradient iteration method to establish an iterative hard thresholding algorithm in their model. The Riemannian optimization is utilized to construct the manifold for low-rank tensors in~\cite{Silva, Kasai, Kressner2014}.

\subsection{Trace Norm Based Methods}
In~\cite{Liu2009} and a subsequent paper~\cite{Liu2013}, the authors generalize matrix completion to study tensors and solve the tensor completion problem by considering the following optimization:
\begin{equation}\begin{split}
    \min_{\mathcal{Y}}&:\Vert\mathcal{Y}\Vert_*,\\
    \text{s.t.}&:\mathcal{Y}_{\Omega}=\mathcal{X}_{\Omega},\\
    \end{split}\label{eq6}
\end{equation}
where $\mathcal{X}\in\mathbb{R}^{n_1\times...\times n_d}$ is the observed tensor, $\mathcal{Y}\in\mathbb{R}^{n_1\times...\times n_d}$ is the estimated tensor, $\Omega$ is the set containing indices of all observed entries in tensor $\mathcal{X}$, and $\Vert\cdot\Vert_*$ is the tensor trace norm. The tensor trace norm is a relaxation of the tensor $n$-rank (rank$_n(\mathcal{X})$, see section~\ref{Decomposition}), and is defined as a convex combination of the trace norms of all unfolding matrices. When the noises are included, the optimization problem is now described by
\begin{equation}
    \begin{split}
    \min_{\mathcal{Y}}\quad & \Vert\mathcal{Y}\Vert_*:=\sum_{k=1}^d \alpha_k\Vert \boldsymbol{Y}_{(k)}\Vert_*\\
    \text{subject to}\quad & \mathcal{Y}_{\Omega}=\mathcal{X}_{\Omega}+\mathcal{E}_{\Omega}\\
    \end{split}
    \label{eq7}
\end{equation}
where the $\alpha_k$'s are non-negative weights satisfying $\sum_{k=1}^d\alpha_k=1$, and $\mathcal{E}_{\Omega}$ is the error. The optimization problem~\eqref{eq7} is called a sum of nuclear norm (SNN) model. Note that we do not impose any data generation assumptions in \eqref{eq7}. If the noise $\mathcal{E}_{\Omega}$ is assumed to be Gaussian, then by considering maximizing the likelihood function under the constraint, the SNN model becomes
\begin{equation}
    \min_{\mathcal{Y}}\frac{\lambda}{2}\Vert \mathcal{P}_{\Omega}(\mathcal{Y}-\mathcal{X})\Vert^2+\sum_{k=1}^d \alpha_k\Vert \boldsymbol{Y}_{(k)}\Vert_*,
    \label{eq8}
\end{equation}
where $\lambda>0$ is a tuning parameter, $\mathcal{P}_{\Omega}(\cdot)$ denotes all the entries in the observed index set $\Omega$, $\Vert\cdot\Vert$ is the tensor norm defined in \eqref{norm}, and $\Vert\cdot\Vert_*$ is the matrix trace norm~\cite{Song2019}. This optimization problem can be solved by block coordinate descent algorithms~\cite{Liu2009} and splitting methods (e.g., Alternating Direction Method of Multipliers, ADMM) \cite{Gandy, Tomioka2010, Signoretto}.

Using a similar model as \eqref{eq6}, Mu et al.~\cite{Mu} propose to apply the trace norm on a balanced unfolding matrix instead of utilizing the summation of trace norms in \eqref{eq7}. In the literature, it is also common to consider alternative norms such as the incoherent trace norm~\cite{Yuan2017} and tensor nuclear norm~\cite{Kilmer, Zhang2014}. There are other studies that impose trace norms on the factorized matrices rather than unfolding matrices~\cite{Liu2014, Ying, Mardani}; these approaches can be viewed as a combination of decomposition based and trace norm based completion methods.

\section{Tensor Regression}\label{TR}
In this section, we review tensor regression methods, where the primary goal is to analyze the association between tensor-valued objects and other variables. Based on the role that the tensor plays in the regression, the problem can be further categorized into tensor predictor regression (with tensor-valued predictors and a univariate or multivariate response variable) and tensor response regression (with tensor-valued response and predictors that can be a vector, a tensor or even multiple tensors).

\subsection{Tensor Predictor Regression}
\begin{figure}[htp]
	{\centering
	\includegraphics[width=.8\linewidth]{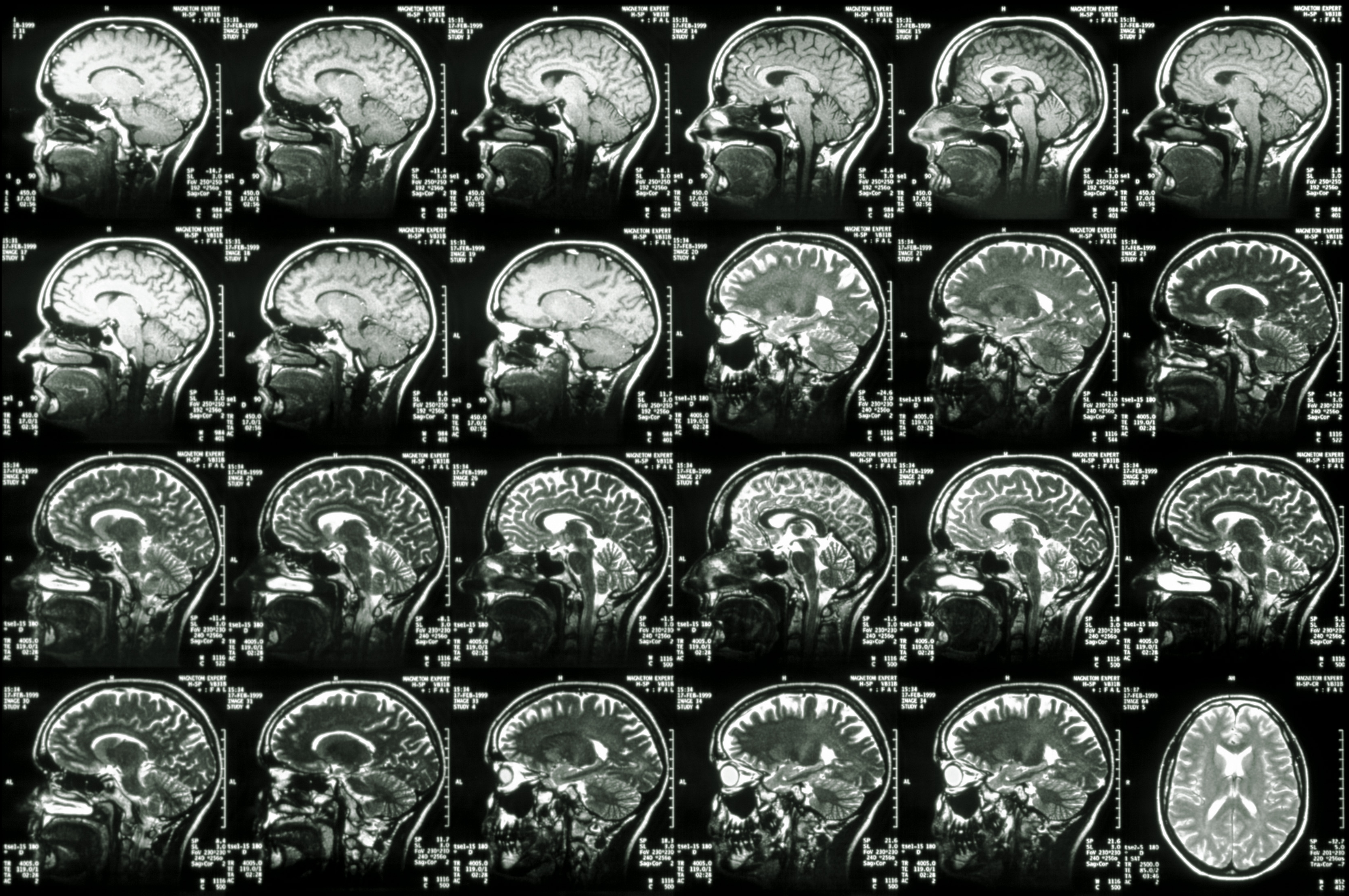}
		\caption{An example of 3D magnetic resonance imaging (MRI). The image is adapted with permissions from Science Photo Library. url: \url{https://www.sciencephoto.com/media/306963/view}}
	\label{f5}}
\end{figure}

Many tensor predictor regression methods are motivated by the need to analyze anatomical magnetic resonance imaging (MRI) data~\cite{Guo, Zhou}. Usually stored in the form of 3D images (see Figure~\ref{f5} for an example), MRI presents the shape, volume, intensity, or developmental changes in brain tissues and blood brain barrier. These characteristics are closely related to the clinical outcomes including diagnostic status, and cognition and memory score. It is hence natural to formulate a tensor predictor regression to model the changes of these scalar or vector-valued clinical outcomes with respect to the tensor-valued MRI images.

In medical imaging analysis, conventional approaches are generally based on vectorized data, either by summarizing the image data through a small number of preidentified regions of interest (ROIs), or by transforming the entire image into a long vector. The former is highly dependent on the prior domain knowledge and does not fully utilize the information in the raw image, and the latter suffers from the high-dimensionality of voxels in the 3D image and abandons important spatial information during the vectorization process. In order to circumvent these limitations, a class of regression methods have been developed to preserve the tensor structure. Specifically, given a univariate response $Y$ (e.g. memory test score, disease status) and a tensor-valued predictor $\mathcal{X}\in\mathbb{R}^{n_1\times...\times n_d}$ (e.g. 3D image), Guo et al.~\cite{Guo} propose a linear regression model
\begin{equation}
    Y=\langle \mathcal{W}, \mathcal{X} \rangle+b,
    \label{eq1}
\end{equation}
where $\langle\cdot,\cdot\rangle$ is the tensor inner product defined in \eqref{inner product}, $\mathcal{W}$ is the coefficient tensor, and $b$ is the error. While model~(\ref{eq1}) is a direct extension of a classical linear regression model, the extension can result in the explosion of the number of unknown parameters. Specifically, the coefficient tensor $\mathcal{W}$  includes $\prod_{i=1}^d n_i$ free parameters, which far exceeds the typical sample size. To address this issue, Guo et al.~\cite{Guo} impose a rank-$r$ CP structure (\ref{CP}) on $\mathcal{W}$, which reduces the number of parameters in $\mathcal{W}$ to $r\sum_{i=1}^d n_i$. 

Li et al.~\cite{Li2016} extend model (\ref{eq1}) to the multivariate response $\boldsymbol{Y}=(Y_1,Y_2,...,Y_q)^{\top}$ case, where each marginal response $Y_k~(1\leq k\leq q)$ is assumed to be the summation of $\langle\mathcal{X}, \mathcal{B}_k\rangle$ and an error term, where $\mathcal{X}$ is the predictor tensor, and $\mathcal{B}_k\in\mathbb{R}^{n_1\times...\times n_d}$ is the coefficient tensor. Under the assumption that the coefficients share common features, the coefficient tensors are further formulated into a stack $\mathcal{B}=[\mathcal{B}_1,...,\mathcal{B}_q]\in\mathbb{R}^{n_1\times...\times n_d\times q}$, on which a CP structure is imposed for parameter number reduction. 

Additionally, Zhou et al.~\cite{Zhou} integrate model~(\ref{eq1}) with the generalized linear regression framework, and incorporate the association between response and other adjusting covariates into the model. Consider a scalar response $Y$, a tensor-valued predictor $\mathcal{X}\in\mathbb{R}^{n_1\times...\times n_d}$ and vectorized covariates $\boldsymbol{z}\in\mathbb{R}^{n_0}$ (e.g., demographic features), the generalized linear model is given by
\begin{equation}
    g\{\mathbb{E}(Y)\}=b+\boldsymbol{\gamma}^{\top}\boldsymbol{z}+\langle\mathcal{W},\mathcal{X}\rangle,
    \label{eq2}
\end{equation}
where $\boldsymbol{\gamma}$ is the vector coefficient for $\boldsymbol{z}$, $g(\cdot)$ is a link function, and $\mathcal{W}$ is the coefficient tensor where a CP structure is assumed. In model \eqref{eq2}, Li et al.~\cite{Li2018} impose a Tucker decomposition on $\mathcal{W}$, and demonstrate that the Tucker structure allows for more flexibility.

In order to accommodate longitudinal correlation of the data in imaging analysis, Zhang et al.~\cite{Zhang2019} extend model \eqref{eq2} in the generalized estimating equation setting and establish asymptotic properties of the method. Hao et al.~\cite{Hao2019} show that the linearity assumption in \eqref{eq1} may be violated in some applications, and propose a nonparametric extension of \eqref{eq1} that accommodates nonlinear interactions between the response and tensor predictor. Zhang et al.~\cite{Zhang2020} use importance sketching to reduce the high computational cost associated with the low-rank factorization in tensor predictor regression, and establish the optimality of their method in terms of reducing mean squared error under the Tucker structure assumption and randomized Gaussian design. Beyond the regression framework, Wimalawarne et al.~\cite{Wimalawarne} propose a binary classification method by considering a logistic loss function and various tensor norms for regularization. 

\subsection{Tensor Response Regression}
\begin{figure}[htp]
	{\centering
	\includegraphics[width=.6\linewidth]{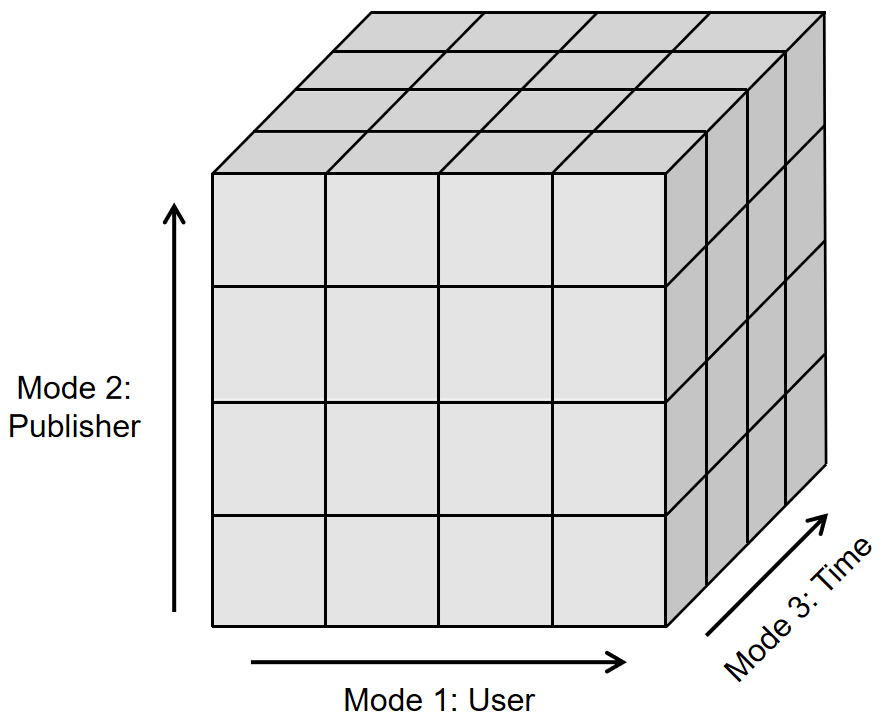}
		\caption{An illustration of click through rate data, which is formulated as a three-mode tensor, where each voxel represents the click-through rate of user $i$ reacting to advertisements from publisher $j$ at time $k$.}
	\label{f6}}
\end{figure}

While the main focus of tensor predictor regression is analyzing the effects of tensors on the response variables, researchers are also interested in studying how tensor-valued outcomes change with respect to covariates. For example, an important question in MRI studies is to compare the scans of brains between subjects with neurological disorders (e.g., attention deficit disorder) and normal controls, after adjusting for other covariates such as age and sex~\cite{Li2016}. This problem can be formulated as a tensor response regression problem where the MRI data, usually taking the form of a three-dimensional image, is the tensor-valued response, and other variables are predictors. Apart from medical imaging analysis, tensor response regression is also useful in the advertisement industry. For example, the click-through rate (CTR) of digital advertisements is often considered to be a significant indicator of the effectiveness of an advertisement campaign. Thus an important business question is to understand how CTR is affected by different features. Since the CTR data can be formulated as a high-dimensional tensor (see Figure~\ref{f6}), we can develop a regression model to address this problem, where the click-through rate on target audience is the tensor-valued response, and the features of advertisements are predictors of interest.

Given a $d$th-order tensor response $\mathcal{Y}\in\mathbb{R}^{n_1\times...\times n_d}$ and a vector predictor $\boldsymbol{x}\in\mathbb{R}^{q}$, Rabusseau and Kadri~\cite{Rabusseau2016} and Sun and Li~\cite{Sun2017} propose a linear regression model
\begin{equation} 
\mathcal{Y}=\mathcal{B}\bar{\times}_{d+1}\boldsymbol{x}+\mathcal{E},
\label{eq3}
\end{equation}
where $\mathcal{B}\in\mathbb{R}^{n_1\times n_2\times...\times n_d\times q}$ is an $(d+1)$th-order tensor coefficient, $\mathcal{E}$ is an error tensor independent of $\boldsymbol{x}$, and $\bar{\times}_{d+1}$ is the $(d+1)$-mode vector product. Without loss of generality, the intercept is set to be zero to simplify the presentation. 

Both studies~\cite{Rabusseau2016, Sun2017} propose to estimate the coefficients $\mathcal{B}$ by solving an optimization problem, which consists of a squared tensor norm of the difference between observed and estimated response $\Vert \mathcal{Y}-\mathcal{B}\bar{\times}_{d+1}\boldsymbol{x}\Vert^2$ and a sparsity structure. In Rabusseau and Kadri~\cite{Rabusseau2016}, the sparsity is achieved by a $L_2$-penalty on parameters. In Sun and Li~\cite{Sun2017}, the sparsity structure is realized through a hard-thresholding constraint on the coefficients. For both studies, decomposition structures are imposed on the tensor coefficient $\mathcal{B}$ to facilitate parsimonious estimation of high-dimensional parameters.

Lock~\cite{Lock} further extends (\ref{eq3}) to a tensor-on-tensor regression model, allowing a predictor of arbitrary order. Given $N$ independent samples, the responses can be stacked into a tensor $\mathcal{Y}\in\mathbb{R}^{N\times m_1\times m_2\times...\times m_q}$, and the predictors are denoted by $\mathcal{X}\in\mathbb{R}^{N\times n_1\times n_2\times...\times n_d}$. Lock~\cite{Lock} proposes the following model:
\begin{equation}
\mathcal{Y}=\mathcal{X}*\mathcal{B}+\mathcal{E},
    \label{eq4}
\end{equation}
where $*$ is the tensor contraction product defined in  \eqref{contraction product}, $\mathcal{B}\in\mathbb{R}^{n_1\times...\times n_d\times m_1\times...\times m_q}$ is the coefficient tensor and $\mathcal{E}$ denotes the error. A CP structure is imposed on $\mathcal{B}$ to achieve parsimony in parameters. The estimation of $\mathcal{B}$ is also transformed into an optimization problem, and a $L_2$-penalty is included in the loss function to prevent over-fitting. Under a similar modeling framework, Gahrooei et al.~\cite{Gahrooei} develop a multiple tensor-on-tensor regression model, where the predictors are a set of tensors with various orders and sizes. 

Based on (\ref{eq4}), Li and Zhang~\cite{Li2017} propose a tensor response regression that utilizes the envelope method to remove redundant information from the response. Raskutti et al.~\cite{Raskutti} analyze the tensor regression problem with convex and weakly decomposable regularizers. In their regression model, both the predictors and the responses can be tensors, and the low-rankness assumption is realized by a nuclear norm penalty. Zhou et al.~\cite{Zhou2020} focus on tensor regression where the response is a partially observed dynamic tensor, and impose low-rankness, sparsity and temporal smoothness constraints in the optimization. Chen et al.~\cite{Chen2019} extend model \eqref{eq4} to the generalized tensor regression setting and utilize a projected gradient descent algorithm to solve the non-convex optimization.

\section{Bayesian Methods in Tensor Completion}\label{BD}
In Section~\ref{DE}, we mention that the tensor completion tasks can be realized by performing decomposition on partially observed tensors and using the inferred decomposition structure to impute the missing data (e.g., the Missing-Skipping methods). Bayesian tensor decomposition methods can be naturally applied to study partially observed tensors.  Generally, a large proportion of Bayesian decomposition methods are based on CP~\eqref{CP} or Tucker decomposition~\eqref{Tucker}. A class of nonparametric methods have also been proposed to model complex non-linear interactions among latent factors. Recently, more decomposition structures are analyzed under the Bayesian framework (e.g., tensor ring decomposition~\cite{Long2020}, tensor train decomposition~\cite{Hu2022} and neural decomposition~\cite{He2018}). A summary of the methods discussed in this section is given in Table~\ref{T1}.

\subsection{Bayesian CP-Based Decomposition}
Under the Bayesian framework, Xiong et al.~\cite{Xiong2010} utilize a CP decomposition based method to model time-evolving relational data in recommender systems. In their study, the observed data are formed into a three-dimensional tensor $\mathcal{R}\in\mathbb{R}^{N\times M\times K}$, where each entry $\mathcal{R}_{ij}^k$ denotes user $i$'s rate on item $j$ given time $k$. A CP structure~(\ref{CP}) is then imposed on $\mathcal{R}$:
\begin{equation} 
\mathcal{R}\approx \sum_{d=1}^D \boldsymbol{U}_{d:}\circ \boldsymbol{V}_{d:}\circ \boldsymbol{T}_{d:}=[\![\boldsymbol{U},\boldsymbol{V},\boldsymbol{T}]\!],
\label{eq9}
\end{equation}
where $\boldsymbol{U}, \boldsymbol{V}, \boldsymbol{T}$ are latent factors corresponding to user, item, and time, respectively; and $\boldsymbol{U}_{d:}, \boldsymbol{V}_{d:}, \boldsymbol{T}_{d:}$ represent the $d$th-row  of $\boldsymbol{U}, \boldsymbol{V}$ and $\boldsymbol{T}$. Xiong et al.~\cite{Xiong2010} assume a Gaussian distribution for the continuous entries $R_{ij}^k$ conditional on $\boldsymbol{U}, \boldsymbol{V}, \boldsymbol{T}$ as follows,  
\begin{equation} 
R_{ij}^k|\boldsymbol{U},\boldsymbol{V},\boldsymbol{T}\sim \mathcal{N}(\langle \boldsymbol{U}_{:i},\boldsymbol{V}_{:j},\boldsymbol{T}_{:k}\rangle,\alpha^{-1}),
\label{eq10}
\end{equation}
where $\alpha$ is the precision, and $\langle \boldsymbol{U}_{:i},\boldsymbol{V}_{:j},\boldsymbol{T}_{:k}\rangle$ is the inner product of three $D$-dimensional vectors defined as
$$\langle \boldsymbol{U}_{:i},\boldsymbol{V}_{:j},\boldsymbol{T}_{:k}\rangle=\sum_{d=1}^D U_{di}V_{dj}T_{dk}.$$
A complete Bayesian setting requires full specification of the parameter priors. In the study, multivariate Gaussian priors are put on the latent vectors corresponding to users and items
\begin{gather}
    \boldsymbol{U}_i\sim \mathcal{N}(\boldsymbol{\mu}_U, \boldsymbol{\Lambda}_U^{-1}), \quad i=1,2,...,N,\label{eq11}\\
    \boldsymbol{V}_j\sim \mathcal{N}(\boldsymbol{\mu}_V, \boldsymbol{\Lambda}_V^{-1}), \quad j=1,2,...,M, \label{eq12}
\end{gather}
and each time feature vector is assumed to depend only on its immediate predecessor due to temporal smoothness:
\begin{align}
   & \boldsymbol{T}_k\sim \mathcal{N}(\boldsymbol{T}_{k-1}, \boldsymbol{\Lambda}_T^{-1}),\quad k=1,2,...,K,\label{eq13}\\
   &  \boldsymbol{T}_0\sim\mathcal{N}(\boldsymbol{\mu}_T, \boldsymbol{\Lambda}_T^{-1}).\label{eq14}
\end{align}

Moreover, Xiong et al.~\cite{Xiong2010} consider a hierarchical Bayesian structure where the hyper-parameters $\boldsymbol{\Theta}_U\equiv\{\boldsymbol{\mu}_U,\boldsymbol{\Lambda}_U\}, \boldsymbol{\Theta}_V\equiv\{\boldsymbol{\mu}_V,\boldsymbol{\Lambda}_V\}, $ and $\boldsymbol{\Theta}_T\equiv\{\boldsymbol{\mu}_T,\boldsymbol{\Lambda}_T\}$ are viewed as random variables, and their prior distributions (i.e., hyper-priors), denoted by $p(\cdot)$, are 
\begin{equation}
    \begin{split}
        p(\boldsymbol{\Theta}_U)&=
        p(\boldsymbol{\mu}_U|\boldsymbol{\Lambda}_U)p(\boldsymbol{\Lambda}_U)=\mathcal{N}(\boldsymbol{\mu}_0,(\beta_0\boldsymbol{\Lambda}_U)^{-1})\mathcal{W}(\boldsymbol{\Lambda}_U|\boldsymbol{W}_0,\nu_0),\\
        p(\boldsymbol{\Theta}_V)&=
        p(\boldsymbol{\mu}_V|\boldsymbol{\Lambda}_V)p(\boldsymbol{\Lambda}_V)=\mathcal{N}(\boldsymbol{\mu}_0,(\beta_0\boldsymbol{\Lambda}_V)^{-1})\mathcal{W}(\boldsymbol{\Lambda}_V|\boldsymbol{W}_0,\nu_0),\\
        p(\boldsymbol{\Theta}_T)&=
        p(\boldsymbol{\mu}_T|\boldsymbol{\Lambda}_T)p(\boldsymbol{\Lambda}_T)=\mathcal{N}(\boldsymbol{\mu}_0,(\beta_0\boldsymbol{\Lambda}_T)^{-1})\mathcal{W}(\boldsymbol{\Lambda}_T|\boldsymbol{W}_0,\nu_0).\\
    \end{split}
    \label{eq15}
\end{equation}
Here $\mathcal{W}(\boldsymbol{\Lambda}|\boldsymbol{W}_0,\nu_0)$ is the Wishart distribution of a $D\times D$ random matrix $\boldsymbol{\Lambda}$ with $\nu_0$ degrees of freedom and a $D\times D$ scale matrix $\boldsymbol{W}_0$:
$$\mathcal{W}(\boldsymbol{\Lambda}|\boldsymbol{W}_0,\nu_0)\propto |\boldsymbol{\Lambda}|^{(\nu_0-D-1)/2} \exp\left(-\frac{\text{Tr}(\boldsymbol{W}_0^{-1}\boldsymbol{\Lambda})}{2}\right).$$

Also, a Wishart prior is put on the precision $\alpha$
\begin{equation}
    p(\alpha)=\mathcal{W}(\alpha|\tilde{W}_0,\tilde{\nu}_0).
    \label{eqn3}
\end{equation}
The priors in~\eqref{eq15}~and~\eqref{eqn3} are conjugate priors for the Gaussian parameters to help simplify the posterior computation. The parameters $\boldsymbol{\mu}_0,\beta_0,\boldsymbol{W}_0,\nu_0,\tilde{W}_0$ and $\tilde{\nu}_0$ can be chosen by prior knowledge or tuned by model training.

The Bayesian model in \eqref{eq10}--\eqref{eq15} is called a Bayesian Probabilistic Tensor Factorization (BPTF). The posterior distribution of the BPTF model is obtained by Markov Chain Monte Carlo (MCMC) with Gibbs sampling~\cite{Gibbs}. While Xiong et al.~\cite{Xiong2010} use the BPTF model to perform tensor decomposition on continuous rating data in recommender systems, similar priors have been adapted in other applications and data types. For example, Chen et al.~\cite{Chen2019b} formulate the spatio-temporal traffic data as a third-order tensor (road segment$\times$day$\times$time of day), where a CP structure is assumed and a Gaussian-Wishart prior is put on the latent factors for conjugacy. A similar model has been used to study multi-relational network~\cite{Gao2012}, where the interaction data form a partially symmetric third-order tensor and the tensor entries are binary indicators of whether a certain type of relationship exists. Correspondingly, a sigmoid function is employed in \eqref{eq10} to map the outer product of latent factors onto the range $[0,1]$.

In addition, Schein et al.~\cite{Schein2014} develop a Poisson tensor factorization (PTF) method to deal with dyadic interaction data in social networks. Specifically, the interaction data are formulated as a fourth-order tensor $\mathcal{X}$, where $\mathcal{X}_{ijat}$ denotes the number of interactions within a discrete time interval $t$ involving a particular sender $i$, receiver $j$, and action-type $a$. A Poisson distribution is employed to connect the CP structure to the count-valued data:
\begin{equation} \mathcal{X}_{ijat}\sim \text{Poisson}(\sum_{k=1}^K \theta_{ik}^s\theta_{jk}^r\psi_{ak}\delta_{tk}).
\label{eq16}
\end{equation}
Here $\theta_{ik}^s, \theta_{jk}^r, \psi_{ak}$ and $\delta_{tk}$ represent the latent factors corresponding to the sender, receiver, action-type and time interval, respectively. Gamma priors are then assigned to the latent factors, 
\begin{equation}\begin{split}
    \theta_{ik}^s&\sim\text{Gamma}(a,b),\\
    \theta_{jk}^r&\sim\text{Gamma}(a,b),\\
    \psi_{ak}&\sim\text{Gamma}(c,d),\\
    \delta_{tk}&\sim\text{Gamma}(e,f).
    \label{eqs1}
\end{split}\end{equation}
Schein et al.~\cite{Schein2014} then represent the Poisson likelihood (\ref{eq16}) as a sum of $K$ independent Poisson random variables, and derive a Variational Bayesian (VB) algorithm to make inference on the posterior distribution.

All the aforementioned methods assume that the interactions among the latent factors are multi-linear, which may not necessarily hold in practice. To address this issue, Liu et al.~\cite{Liu2018} consider a neural CP decomposition that exploits both neural networks and probabilistic methods to capture potential nonlinear interactions among the tensor entries. Given a tensor $\mathcal{X}$ and the latent matrices in its CP structure $\boldsymbol{U}^{1},...,\boldsymbol{U}^D$, the distribution of $\mathcal{X}$ conditional on $\boldsymbol{U}^{1},...,\boldsymbol{U}^D$ is given by
$$ p(\mathcal{X}|\{\boldsymbol{U}^d\}_{d=1}^D)=\prod_{i_1,...,i_D}\mathcal{N}(x_{i_1\ldots i_D}|\mu(\boldsymbol{u}_{i_1\ldots i_D}),\sigma^2(\boldsymbol{u}_{i_1\ldots i_D})),$$
where $\boldsymbol{u}_{i_1\ldots i_D}=(U_{i_1:}^1,...,U_{i_D:}^D)\in\mathbb{R}^{DR}$ is a long vector generated by concatenating the elements in the $i_d$th row of the factor matrix $U^d$. In order to accommodate nonlinear interactions between latent factors, $\mu$ and $\sigma^2$ are defined as functions of $\boldsymbol{u}_{i_1\ldots i_D}$ ($\mu=\mu(\boldsymbol{u}_{i_1\ldots i_D}), \sigma^2=\sigma^2(\boldsymbol{u}_{i_1\ldots i_D})$). In particular, the two functions $\mu(\cdot)$ and $\sigma^2(\cdot)$ are modeled by two neural networks with the same input $\boldsymbol{u}_{i_1\ldots i_D}$,
\begin{equation*}
    \begin{split}
        \mu &=\boldsymbol{w}_\mu^{\top} \boldsymbol{h}(\boldsymbol{u}_{i_1\ldots i_D})+b_\mu,\\
        \log\sigma^2 &=\boldsymbol{w}_{\sigma}^{\top}\boldsymbol{h}(\boldsymbol{u}_{i_1\ldots i_D})+b_{\sigma}, 
    \end{split}
\end{equation*}
where $\boldsymbol{h}(\boldsymbol{u}_{i_1\ldots i_D})$ is a nonlinear hidden layer shared by these two neural networks, and is defined as a {\it tanh} activation function in \cite{Liu2018}:
$$ \boldsymbol{h}(\boldsymbol{u}_{i_1\ldots i_D})=tanh(\boldsymbol{W}^{\top}\boldsymbol{u}_{i_1\ldots i_D}+\boldsymbol{b}).$$

As discussed in Section~\ref{Decomposition}, determining the rank of CP can be challenging in practice. Even for a noise-free tensor, its rank specification is an NP-hard problem \cite{Hastad}. In order to determine the CP rank, a common practice is to fit models with different ranks and choose the best rank based on certain criteria. Nevertheless, this approach may suffer from a low stability issue and a high computational cost. An alternative approach is to use sparsity-inducing priors. For example, in~\cite{Rai2014} and a subsequent work~\cite{Rai2015}, the authors propose a Bayesian low-rank CP decomposition method, which utilizes the multiplicative gamma process (MGP) prior~\cite{MGP} to automatically infer the rank. Specifically, given a CP structure
$$ \mathcal{X}=\sum_{r=1}^R \lambda_r\cdot\boldsymbol{u}_r^{(1)}\circ\boldsymbol{u}_r^{(2)}\circ\cdots\circ\boldsymbol{u}_r^{(K)},$$
the following priors are put on the vector $\boldsymbol{\lambda}=(\lambda_1,\lambda_2,...,\lambda_R)$:
\begin{align} 
& \lambda_r\sim \mathcal{N}(0, \tau_r^{-1}),\quad 1\leq r\leq R \label{eqs2}\\
& \tau_r=\prod_{l=1}^r\delta_l,\quad \delta_l\sim \text{Gamma}(a_c,1),\quad a_c>1. \label{eqs3}
\end{align}
In MGP prior, as $r$ increases, the precision $\tau_r$ takes large values hence shrinks  $\lambda_r$ towards zero. Small $\lambda_r$ values indicate that the term $\lambda_r\cdot\boldsymbol{u}_r^{(1)}\circ\boldsymbol{u}_r^{(2)}\circ\cdots\circ\boldsymbol{u}_r^{(K)}$ does not have a significant impact on the CP structure, hence could be removed from the model. Two generalizations of MGP prior are further developed, including truncation based variant MGP-CP$^t$ and the adaptive variant MGP-CP$^a$,  to automatically infer the rank $R$~\cite{Rai2014, Rai2015}.

Hu et al.~\cite{Hu2015} develop a Bayesian non-negative tensor factorization that deals with count-valued data and automatically infers the rank of CP decomposition. In their work, the Poisson distribution is utilized to establish a connection between the CP structure and the count-valued data. Given a tensor $\mathcal{Y}\in\mathbb{R}^{n_1\times...\times n_K}$ and its entries $\boldsymbol{i}=\{i_1,...,i_K\}$, we have
$$ \mathcal{Y}_{\boldsymbol{i}}\sim \text{Poisson}\left(\sum_{r=1}^R \lambda_r\prod_{k=1}^K u_{i_kr}^{(k)}\right).$$
The non-negativity constraints on the factor matrices $\boldsymbol{U}^{(1)},...,\boldsymbol{U}^{(K)}$ ($\boldsymbol{U}^{(k)}=[\boldsymbol{u}_1^{(k)},...,\boldsymbol{u}_R^{(k)}], k=1,2,...,K$) are naturally satisfied by imposing Dirichlet priors on the factors $\boldsymbol{u}_r^{(k)}=[u_{1r}^{(k)},...,u_{i_kr}^{(k)}]^\top$:
$$ \boldsymbol{u}_r^{(k)}\sim \text{Dir}(a^{(k)},...,a^{(k)}),$$
and a gamma-beta hierarchical prior is put on $\lambda_r$ to promote the automatic rank specification:
\begin{gather} 
\lambda_r\sim\text{Gamma}(g_r,\frac{p_r}{1-p_r}),\label{eqs4}\\
p_r\sim\text{Beta}(c\epsilon, c(1-\epsilon))~~~\text{for some}~c > 0.\label{eqs5}
\end{gather}

Similar to the MGP prior in~\eqref{eqs2} and \eqref{eqs3}, the gamma-beta hierarchical prior in \eqref{eqs4} and \eqref{eqs5} also shrinks $\lambda_r$ to zero as $r$ increases, and is thus able to select the CP rank. This model is also extended to binary data by adding an additional layer $b_{\boldsymbol{i}}=\boldsymbol{1}(y_{\boldsymbol{i}}\geq 1)$, which takes a count-valued entry $y_{\boldsymbol{i}}$ in $\mathcal{Y}$ and thresholds this latent count at one to generate binary-valued entries $b_{\boldsymbol{i}}$~\cite{Hu2015b}. 

Instead of imposing sparsity priors on the core elements of CP structure, Zhao et al.~\cite{Zhao2015} place a hierarchical prior over the latent factors. Let $\mathcal{X}\in\mathbb{R}^{I_1\times\cdots\times I_N}$ have a CP structure
$$ \mathcal{X}=[\![\boldsymbol{A}^{(1)},...,\boldsymbol{A}^{(N)}]\!],$$
where $\boldsymbol{A}^{(n)}=[\boldsymbol{a}_1^{(n)},...,\boldsymbol{a}_{I_n}^{(n)}]$ $(n=1,2,...,N)$ are latent factors. Let $\boldsymbol{\lambda}=[\lambda_1,...,\lambda_R]$ and $\boldsymbol{\Lambda}=$ diag($\boldsymbol{\lambda}$). The prior distribution of $\boldsymbol{A}^{(n)}$ is 
$$ p(\boldsymbol{A}^{(n)}|\boldsymbol{\lambda})=\prod_{i_n=1}^{I_n}\mathcal{N}(\boldsymbol{a}_{i_n}^{(n)}|\boldsymbol{0},\boldsymbol{\Lambda}^{-1}),\quad n=1,2,\ldots,N.$$ A hyperprior is further defined over $\boldsymbol{\lambda}$, which is factorized over the latent dimensions
$$ p(\boldsymbol{\lambda})=\prod_{r=1}^R \text{Gamma}(\lambda_r|c_0^r,d_0^r).$$
Here $R$ is a pre-specified maximum possible rank. The latent vectors (the $r$th row of all latent matrices) will shrink to a zero vector as $\lambda_r^{-1}$'s approach to zero. This model can also accommodate various types of outliers and non-Gaussian noise through the introduction of a sparsity structure, and the tradeoff between the low-rankness approximation and the sparse representation can be learned automatically by maximizing the model evidence~\cite{Zhao2015b}.

In real-world applications including recommender systems, image/video data analysis and internet networks, the data are sometimes produced continuously (i.e., streaming data). Therefore it is of interest to generalize the tensor decomposition models to analyze such data in a real time manner, where the model parameters can be updated efficiently upon receiving new data without retrieving previous entries. To this end, a class of streaming tensor decomposition methods have been developed, and some are analyzed under the Bayesian CP framework~\cite{Zhang2018, Du2018, Fang2021}. In general, these algorithms start with a prior distribution of unknown parameters and then infer a posterior that best approximates the joint distribution of these parameters upon the arrival of new streaming data. The estimated posterior is then used as the prior for the next update. These methods are implemented either by streaming variational Bayes (SVB)~\cite{Zhang2018, Du2018}, or assume-density filtering (ADF) and expectation-propagation (EP)~\cite{Fang2021}.

\subsection{Tucker-based Bayesian Decomposition Methods}
Compared to the CP decomposition, the Tucker structure~(\ref{Tucker}) can model more complex interactions between latent factors. One of the early works that employs a probabilistic Tucker structure is proposed by Chu and Ghahramani~\cite{Chu2009}, where a probabilistic framework called pTucker is developed to perform a decomposition on partially observed tensors. Given a continuous third-order tensor $\mathcal{Y}\in\mathbb{R}^{n\times m\times d}$, a Gaussian distribution is assigned to each entry of the tensor $\mathcal{Y}$, 
$$ \mathcal{Y}_{ijr}|\mathcal{T}\sim \mathcal{N}(\mathcal{F}_{ijr}, \sigma^2).$$
Here $\mathcal{F}$ has a Tucker structure with a core tensor $\mathcal{T}$
$$ \mathcal{F}_{ijr}=\text{vec}(\mathcal{T})^{\top} (\boldsymbol{v}_r\otimes\boldsymbol{z}_j\otimes\boldsymbol{x}_i),$$
where $\otimes$ is the Kronecker product, and $\boldsymbol{v}_r, \boldsymbol{z}_j$ and $\boldsymbol{x}_i$ are latent vectors. Next, independent standard normal distributions are specified over the entries in $\mathcal{T}$ as priors:
$$ \mathcal{T}_{kls}\sim\mathcal{N}(0,1),\quad\forall k,l,s.$$

By integrating out the core tensor $\mathcal{T}$ from the joint distribution $\prod_{i,j,r}p(\mathcal{Y}_{ijr}|\mathcal{T})\prod_{k,l,s}p(\mathcal{T}_{kls})$, the observational array still follows a Gaussian distribution:
$$ \text{vec}(\mathcal{Y})\sim \mathcal{N}(\boldsymbol{0}, \boldsymbol{U}\boldsymbol{U}^{\top}+\sigma^2\boldsymbol{I}),$$
where $\text{vec}(\mathcal{Y})$ is the vectorized tensor,  $\sigma^2$ is the noise level, and $\boldsymbol{U}=\boldsymbol{V}\otimes\boldsymbol{Z}\otimes\boldsymbol{X},$ where $\boldsymbol{V}, \boldsymbol{Z}$ and $\boldsymbol{X}$ are latent matrices. To complete the Bayesian framework, standard normal distributions are further used as priors for latent components $\boldsymbol{X}, \boldsymbol{Z}$ and $\boldsymbol{V}$. Finally, the latent factors are estimated by maximum a posteriori (MAP) method with gradient descent.

While the MAP method provides an efficient alternative to perform point estimation for latent factors, it also has significant disadvantages including vulnerability to overfitting and incapability of quantifying parameter uncertainties. To this end, various approaches seek to provide a fully Bayesian treatment through inferring the posterior distribution of parameters. For instance, Hayashi et al.~\cite{Hayashi2010} utilize the expectation maximization (EM) method that combines the Laplace approximation and the Gaussian process to perform posterior inference on latent factors. They use the exponential family distributions to connect the Tucker structure with the observed tensor, thus developing a decomposition method that is compatible with various data types. In addition, Schein et al.~\cite{Schein2016} propose a Bayesian Poisson Tucker decomposition (BPTD) that uses MCMC with Gibbs sampling for posterior inference. That method mainly focus on modeling count-valued tensors by putting Poisson priors on the Tucker structure entries and Gamma priors on the latent factors. More recently, Fang et al.~\cite{Fang2021T} develop a Bayesian streaming sparse Tucker decomposition (BASS-Tucker) method to deal with streaming data. BASS-Tucker assigns a spike-and-slab prior over entries of core tensor and employs an extended assumed density filtering (ADF) framework for posterior inference.

Similar to CP-based methods, an important task for Tucker decomposition based methods is to choose an appropriate tensor rank. Unfortunately, this problem is challenging especially when dealing with partially observed data corrupted with noise. Zhao et al.~\cite{Zhao2015T} employ hierarchical sparsity-inducing priors to perform automatic rank determination in their Bayesian tensor decomposition (BTD) model. Specifically, the observed tensor $\mathcal{Y}\in\mathbb{R}^{I_1\times...\times I_N}$ is assumed to follow a Gaussian distribution with the mean following a Tucker structure:
$$ \text{vec}(\mathcal{Y})|\{\boldsymbol{U}^{(n)}\},\mathcal{G},\tau\sim \mathcal{N}((\bigotimes_n\boldsymbol{U}^{(n)}))\text{vec}(\mathcal{G}),\tau^{-1}\boldsymbol{I}),$$
where $\{\boldsymbol{U}^{(n)}\}$ are latent matrices, $\mathcal{G}$ is the core tensor, and $\tau$ is the precision. To allow a fully Bayesian treatment, hierarchical priors are placed over all model parameters. First, a noninformative Gamma prior is assigned to the precision parameter $\tau$
$$ \tau\sim \text{Gamma}(a_0^{\tau}, b_0^{\tau}).$$
Next, a group sparsity prior is employed over the factor matrices, i.e., each $\boldsymbol{U}^{(n)}=[\boldsymbol{u}_{1}^{(n)},...,\boldsymbol{u}_{I_n}^{(n)}]^\top$ ($\boldsymbol{u}_{i_n}^{(n)}$ are latent vectors) is governed by hyper-parameters $\boldsymbol{\lambda}^{(n)}=(\lambda_{1}^{(n)},...,\lambda_{R_n}^{(n)})$, where $\lambda_{r_n}^{(n)}$ controls the precision related to group $r_n$ (i.e., $r_n$th column of $\boldsymbol{U}^{(n)}$). Let $\boldsymbol{\Lambda}^{(n)}=$diag($\boldsymbol{\lambda}^{(n)}$), then the group sparsity prior is given by
$$ \boldsymbol{u}_{i_n}^{(n)}|\boldsymbol{\lambda}^{(n)}\sim \mathcal{N}(\boldsymbol{0}, \boldsymbol{\Lambda}^{(n)^{-1}}),\quad \forall n, \forall i_n.$$

The sparsity assumption is also imposed on the core tensor $\mathcal{G}$. Considering the connection between latent factors and the corresponding entries of the core tensor, the precision parameter for $\mathcal{G}_{r_1,...,r_N}$ can be viewed as the product of precisions over $\{\boldsymbol{u}_{\cdot r_n}^{(n)}\}_{n=1}^N$, which is represented by
$$ \mathcal{G}_{r_1...r_N}|\{\boldsymbol{\lambda}^{(n)}\},\beta\sim\mathcal{N}(0, (\beta\prod_n \lambda_{r_n}^{(n)})^{-1}),$$
or equivalently,
$$ \text{vec}(\mathcal{G})|\{\boldsymbol{\lambda}^{(n)}\},\beta\sim \mathcal{N}(\boldsymbol{0}, (\beta\bigotimes_n\boldsymbol{\Lambda}^{(n)})^{-1}),$$
where $\beta$ is a scaling parameter on which a Gamma prior is placed
$$ \beta\sim \text{Gamma}(a_0^{\beta}, b_0^{\beta}).$$
The hyperprior for $\boldsymbol{\lambda}^{(n)}$ plays a key role for different sparsity-inducing priors. Two options (student-$t$ and Laplace) are commonly used to achieve group sparsity:
$$ \text{Student-}t: \lambda_{r_n}^{(n)}\sim \text{Gamma}(a_0^{\lambda}, b_0^{\lambda}),\quad \forall n,\forall r_n;$$
\begin{align*} \text{Laplace} :&   ~\lambda_{r_n}^{(n)}\sim \text{IG}(1, \frac{\gamma}{2}),\quad \forall n,\forall r_n,\\
& ~\gamma\sim \text{Gamma}(a_0^{\gamma}, b_0^{\gamma}).
\end{align*}

\begin{table*}[htp]
    {\centering
	\caption{Summary of Bayesian tensor decomposition methods.}}\label{T1}
	   \hspace{0.0675\linewidth}\begin{minipage}{0.865\linewidth}
 		\begin{tabular}{|c|c|ccc|}\hline
 		    \multirow{2}{*}{Name} & Decomposition & \multirow{2}{*}{Rank Specification} & Posterior & \multirow{2}{*}{Data Type}\\
 		    & Structure &  & Inference & \\\hline
 		    BPTF~\cite{Xiong2010} &  & Pre-specify & Gibbs & Continuous\\
 		    PLTF~\cite{Gao2012} &  & Pre-specify & Gibbs & Binary\\
 		    BGCP~\cite{Chen2019b} &  & Pre-specify & Gibbs & Continuous\\
 		    PTF~\cite{Schein2014} &  & Pre-specify & VB & Count\\
 		    NeuralCP~\cite{Liu2018} &  & Pre-specify & AEVB & Continuous\\
 		    MGP-CP~\cite{Rai2014} &  & Automatically inferred & Gibbs & Continuous/Binary\\
 		    PGCP~\cite{Rai2015} & CP & Automatically inferred & Gibbs/EM & Binary/Count\\
 		    BNBCP~\cite{Hu2015} & Decomposition & Automatically inferred & Gibbs/VB & Count\\
 		    ZTP-CP~\cite{Hu2015b} &  & Automatically inferred & Gibbs & Binary\\
 		    FBCP~\cite{Zhao2015} &  & Automatically inferred & VB & Continuous\\ 
 		    BRTF~\cite{Zhao2015b} &  & Automatically inferred & VB & Continuous\\
 		    POST~\cite{Du2018} &  & Pre-specify & SVB & Continuous/Binary\\
 		    BRST~\cite{Zhang2018} &  & Automatically inferred & SVB & Continuous\\
 		    SBDT~\cite{Fang2021} &  & Pre-specify & ADF\&EP & Continuous/Binary\\\hline
 		    pTucker~\cite{Chu2009} &  & Pre-specify & MAP/EM & Continuous\\
 		    Hayashi et al.~\cite{Hayashi2010} & \multirow{2}{*}{Tucker} & Pre-specify & EM & All\\
 		    BPTD~\cite{Schein2016} & \multirow{2}{*}{Decomposition} & Pre-specify & Gibbs & Count\\
 		    BTD~\cite{Zhao2015T} &  & Automatically inferred & VB & Continuous\\
 		    BASS-Tucker~\cite{Fang2021T} &  & Pre-specify & ADF\&EP & Continuous\\\hline
 		    InfTucker~\cite{Xu2012} & \multirow{9}{*}{Nonparametric} & \multirow{9}{*}{Pre-Specify} & VEM & \multirow{9}{*}{Binary/Continuous}\\
 		    Zhe et al.~\cite{Zhe2015N} &  &  & VEM & \\
 		    DinTucker~\cite{Zhe2016N} &  &  & VEM & \\
 		    Zhe et al.~\cite{Zhe2016b} &  &  & VI & \\
 		    SNBTD~\cite{Pan2020} &  &  & ADF\&EP & \\
 		    POND~\cite{Tillinghast2020} &  &  & VB & \\
 		    Zhe and Du~\cite{Zhe2018} &  &  & VEM & \\
 		    Wang et al.~\cite{Wang2020} &  &  & VI & \\
 		    BCTT~\cite{Fang2022} &  &  & EP & \\\hline
 		    TR-VBI~\cite{Long2020} & Tensor Ring & Automatically inferred & VB & Continuous\\
 		    KFT~\cite{Hu2022} & Tensor Train & N/A & VI & Continuous\\
 		    He et al.~\cite{He2018} & Neural & N/A & AEVB & All\\\hline
 		\end{tabular}
 	ADF: Assume-density filtering~\cite{ADF}. AEVB: Auto-Encoding Variational Bayes~\cite{AEVB}.
    EM: Expectation maximization. EP: Expectation propagation~\cite{EP}.
    Gibbs: Markov chain Monte Carlo (MCMC) with Gibbs sampling. MAP: Maximum a posteriori.
    SVB: Steaming variational Bayes.
    VB: Variational Bayes. VEM: Variational expectation maximization.
    VI: Variational Inference. N/A: Not applicable. Neural: Neural tensor decomposition.\\
    \end{minipage}
\end{table*}

\subsection{Nonparametric Bayesian Decomposition Methods}
In addition to the aforementioned linear models,  a class of nonparametric Bayesian approaches have been developed to capture the potential nonlinear relationship between tensor entries. One of the pioneering works is InfTucker proposed by Xu et al.~\cite{Xu2012}. Generally, InfTucker maps the latent factors onto an infinite feature space and then performs Tucker decomposition with the core tensor of an infinite size. Let $\mathcal{M}\in\mathbb{R}^{m_1\times...\times m_K}$ be a tensor following a Tucker structure with a core tensor $\mathcal{W}$ and latent factors $\boldsymbol{U}^{(1)},...,\boldsymbol{U}^{(K)}$. One can assign an element-wise standard Gaussian prior over the core tensor $\mathcal{W}$ (vec$(\mathcal{W})\sim \mathcal{N}(\text{vec}(\mathcal{W}); \boldsymbol{0}, \boldsymbol{I})$) and marginalize out $\mathcal{W}$. The marginal distribution of tensor $\mathcal{M}$ is then given by
\begin{equation} 
p(\mathcal{M}|\boldsymbol{U}^{(1)},...,\boldsymbol{U}^{(K)})=\mathcal{N}(\text{vec}(\mathcal{M});\boldsymbol{0},\boldsymbol{\Sigma}^{(1)}\otimes...\otimes \boldsymbol{\Sigma}^{(K)})),
\label{eq17}
\end{equation}
where $\boldsymbol{\Sigma}^{(K)}=\boldsymbol{U}^{(K)}\boldsymbol{U}^{(K)^{\top}}$. Since the goal is to capture the nonlinear relationships, each row $\boldsymbol{u}_t^k$ of the latent factors $\boldsymbol{U}^{(k)}$ is replaced by a nonlinear map $\phi(\boldsymbol{u}_t^k)$. Then a nonlinear covariance matrix $\boldsymbol{\Sigma}^{(k)}=k(\boldsymbol{U}^{(k)}, \boldsymbol{U}^{(k)})$ can be obtained, where $k(\cdot,\cdot)$ is a nonlinear covariance kernel function. In InfTucker~\cite{Xu2012}, $k(\cdot,\cdot)$ is chosen as the radial basis function kernel. After feature mapping, the core tensor $\mathcal{W}$ has the size of the mapped feature vector $\boldsymbol{u}_t^k$ on mode $k$, which can be potentially infinity. Because the covariance of vec($\mathcal{M}$) is a function of the latent factors $\mathcal{U}=\{\boldsymbol{U}^{(1)},...,\boldsymbol{U}^{(K)}\}$, equation  \eqref{eq17}  actually defines a Gaussian process (GP) on tensor entries, where the input is based on the corresponding latent factors $\mathcal{U}$. To encourage sparse estimation, element-wise Laplace priors are assigned on $\mathcal{U}$:
\begin{equation} \boldsymbol{u}_i^{(k)}\sim \mathcal{L}(\lambda)\propto \text{exp}(-\lambda\Vert\boldsymbol{u}_i^{(k)}\Vert_1).
\label{eqn1}\end{equation}

Finally, the observed tensor $\mathcal{Y}$ is sampled from a noisy model $p(\mathcal{Y}|\mathcal{M})$, of which the form depends on the data type of $\mathcal{Y}$. The joint distribution is then given by
$$ p(\mathcal{Y}, \mathcal{M}, \mathcal{U})=p(\mathcal{U})p(\mathcal{M}|\mathcal{U})p(\mathcal{Y}|\mathcal{M}),$$
where $p(\mathcal{U})$ is given by  \eqref{eqn1}, and $p(\mathcal{M}|\mathcal{U})$ is given by \eqref{eq17} with $\boldsymbol{\Sigma}^{(k)}=k(\boldsymbol{U}^{(k)}, \boldsymbol{U}^{(k)})$.

Under a similar modeling framework, Zhe et al.~\cite{Zhe2015N} make two modifications to InfTucker. One is to assign a Dirichlet process mixture (DPM) prior~\cite{DPM} over the latent factors that allows a random number of latent clusters. The other is to utilize a local GP assumption instead of a global GP when generating the observed array given the latent factors, which enables fast computation over subarrays. Specifically, the local GP-based construction is realized by first breaking the whole array $\mathcal{Y}$ into smaller subarrays $\{\mathcal{Y}_1,..,\mathcal{Y}_N\}$. Then for each subarray $\mathcal{Y}_n$, a latent real-valued subarray $\mathcal{M}_n$ is generated by a local GP based on the corresponding subset of latent factors $\mathcal{U}_n=\{\boldsymbol{U}_n^{(1)},...,\boldsymbol{U}_n^{(K)}\}$, and the noisy observation $\mathcal{Y}_n$ is sampled according to $\mathcal{M}_n$, 
\begin{equation*}
p(\mathcal{Y}_n,\mathcal{M}_n|\mathcal{U})=p(\mathcal{M}_n|\mathcal{U}_n)p(\mathcal{Y}_n|\mathcal{M}_n)
=\mathcal{N}(\text{vec}(\mathcal{M}_n); \boldsymbol{0},\boldsymbol{\Sigma}_n^{(1)}\otimes...\otimes\boldsymbol{\Sigma}_n^{(K)})p(\mathcal{Y}_n|\mathcal{M}_n),
\end{equation*}
where $\boldsymbol{\Sigma}_n^{(k)}=k(\boldsymbol{U}_n^{(k)}, \boldsymbol{U}_n^{(k)})$ is the $k$th mode covariance matrix over the sub-factors $\mathcal{U}_n$.

Likewise, DinTucker~\cite{Zhe2016N} consider a local GP assumption and sample each of the subarrays $\{\mathcal{Y}_1,...,\mathcal{Y}_n\}$ from a GP based on the latent factors $\tilde{\mathcal{U}}_n=\{\tilde{\boldsymbol{U}}_n^{(1)},...,\tilde{\boldsymbol{U}}_n^{(K)}\}$. Different from Zhe et al.~\cite{Zhe2015N}, in DinTucker these latent factors are then tied to a set of common latent factors $\mathcal{U}=\{\boldsymbol{U}^{(1)},...,\boldsymbol{U}^{(K)}\}$ via a prior distribution
$$ p(\tilde{\mathcal{U}}_n|\mathcal{U})=\prod_{k=1}^K \mathcal{N}(\text{vec}(\tilde{\boldsymbol{U}}_n^{(k)})|\text{vec}(\boldsymbol{U}^{(k)}), \lambda\boldsymbol{I}),$$
where $\lambda$ is the variance parameter that controls the similarity between $\mathcal{U}$ and $\tilde{\mathcal{U}}_n$. Furthermore, DinTucker divides each subarray $\mathcal{Y}_n$ into $T_n$ smaller subarrays $\mathcal{Y}_n=\{\mathcal{Y}_{n1},...,\mathcal{Y}_{nT_n}\}$ that share the same latent factors $\{\tilde{\mathcal{U}}_n\}$, and their joint probability is given by
\begin{equation*}
p(\mathcal{U}, \{\tilde{\mathcal{U}}_n, \mathcal{M}_n,\mathcal{Y}_n\}_{n=1}^N)
=\prod_{n=1}^N p(\tilde{\mathcal{U}}_n|\mathcal{U})\prod_{t=1}^{T_n}p(\mathcal{M}_{nt}|\tilde{\mathcal{U}}_n)p(\mathcal{Y}_{nt}|\mathcal{M}_{nt}),
\end{equation*}
where $\mathcal{M}_{nt}$ is a latent subarray, and $\mathcal{M}_n=\{\mathcal{M}_{nt}\}_{t=1}^{T_n}$. The local terms require less memory and have a faster processing time than the global term. More importantly, the additive nature of these local terms in the log domain enables distributed inference, which is then realized through the MapReduce system.

While Zhe et al.~\cite{Zhe2015N} and DinTucker~\cite{Zhe2016N} improve the scalability of their GP-based approaches through modeling the subtensors, their methods can still run into challenges when the sparsity level is very high in observed tensors. To address this issue, a class of methods that do not rely on the Kronecker-product structure in the variance \eqref{eq17} are proposed based on the idea of selecting an arbitrary subset of tensor entries for training. Assume that the decomposition is performed on a sparsely observed tensor $\mathcal{Y}\in\mathbb{R}^{d_1\times...\times d_K}$. For each tensor entry $\boldsymbol{i}=(i_1,...,i_K)$, Zhe et al.~\cite{Zhe2016b} first construct an input $\boldsymbol{x_i}$ by concatenating the corresponding latent factors from all the modes: $\boldsymbol{x_i}=[\boldsymbol{u}_{i_1}^{(1)},...,\boldsymbol{u}_{i_K}^{(K)}]$, where $\boldsymbol{u}_{i_k}^{(k)}$ is the $i_k$th row in the latent factor matrix $\boldsymbol{U}^{(k)}$ for mode $k$. Then each $\boldsymbol{x}_{\boldsymbol{i}}$ is transformed to a scalar $m_{\boldsymbol{i}}$ through an underlying function $f: \mathbb{R}^{\sum_{j=1}^K d_j}\to\mathbb{R}$ such that $m_{\boldsymbol{i}}=f(\boldsymbol{x_i})=f([\boldsymbol{u}_{i_1}^{(1)},...,\boldsymbol{u}_{i_K}^{(K)}])$. After that, a GP prior is assigned over $f$ to learn the unknown function: for any set of tensor entries $S=\{\boldsymbol{i}_1,...,\boldsymbol{i}_N\}$, the function values $\boldsymbol{f}_S=\{f(\boldsymbol{x}_{\boldsymbol{i}_1}), ...,f(\boldsymbol{x}_{\boldsymbol{i}_N})\}$ are distributed according to a multivariate Gaussian distribution with mean $\boldsymbol{0}$ and the covariance determined by $\boldsymbol{X}_S=\{\boldsymbol{x}_{\boldsymbol{i}_1},...,\boldsymbol{x}_{\boldsymbol{i}_N}\}$:
\begin{equation} 
p(\boldsymbol{f}_S|\mathcal{U})=\mathcal{N}(\boldsymbol{f}_S|\boldsymbol{0},k(\boldsymbol{X}_S, \boldsymbol{X}_S)),
\label{eq18}
\end{equation}
where $\mathcal{U}$ is the latent factor, and $k(\cdot,\cdot)$ is a nonlinear covariance kernel. Note that this method is equivalent to InfTucker~\cite{Xu2012} if all entries are selected and a Kronecker-product structure is applied in the full covariance. A standard normal prior is assigned over the latent factors, and the observed entries $\boldsymbol{y}=[y_{\boldsymbol{i}_1},...,y_{\boldsymbol{i}_N}]$ are sampled from a model $p(\boldsymbol{y}|\boldsymbol{m})$, where $p(\cdot)$ is selected based on the data type.

Following the sparse GP framework (\ref{eq18}), Pan et al.~\cite{Pan2020} propose the Streaming Nonlinear Bayesian Tensor Decomposition (SNBTD) that performs fast posterior updates upon receiving new tensor entries. Their model is augmented with feature weights to incorporate a linear structure, and the assumed-density-filtering (ADF) framework is extended to perform reliable streaming inference. Also based on (\ref{eq18}), Tillinghast et al.~\cite{Tillinghast2020} utilize convolutional neural networks to construct a deep kernel $k(\cdot,\cdot)$ for GP modeling, which is more powerful in estimating arbitrarily complicated relationships in data compared to the methods based on shallow kernel functions (e.g., RBF kernel).

In some applications, the tensor data are observed with additional temporal information. Various approaches have been proposed to preserve the accurate timestamps and take full advantage of the temporal information. Among these methods, Zhe and Du~\cite{Zhe2018} and Wang et al.~\cite{Wang2020} perform decomposition based on event-tensors to capture complete temporal information, and Fang et al.~\cite{Fang2022} model the core tensor as a time-varying function, where GP prior is placed to estimate different types of temporal dynamics.

\section{Bayesian Methods in Tensor Regression}\label{BR}
Similar to the frequentist tensor regression methods discussed in Section~\ref{TR}, Bayesian tensor regression methods can be categorized into Bayesian tensor predictor regression and Bayesian tensor response regression. We discuss these two classes of methods in Section~\ref{BTPR} and~\ref{BTRR}, and their theoretical properties in Section~\ref{Theo}. We also review posterior computing in Section~\ref{sec:64}. A summary of the methods discussed in this section is given in Table~\ref{T2}.

\begin{table*}[htp]
    {\centering
	\caption{Summary of Bayesian tensor regression methods.}\label{T2}}
	\hspace{0.055\linewidth}\begin{minipage}{0.89\linewidth}
        \begin{tabular}{|c|cccc|}\hline
 		    \multirow{2}{*}{Name} & Predictor & Response & Tensor & \multirow{2}{*}{Algorithm}\\
 		    & Type & Type & Structure & \\\hline
 		    Suzuki~\cite{Suzuki2015} & Tensor & Scalar & CP & Gibbs\\
 		    BTR~\cite{Guhaniyogi2017} & Tensor+Vector & Scalar & CP & Gibbs\\
 		    Zhao et al.~\cite{Zhao2014P} & Tensor & Scalar & Nonparametric & MAP\\
 		    OLGP~\cite{Hou2015} & Tensor & Scalar & Nonparametric & OLGP\\
 		    AMNR~\cite{Imaizumi2016} & Tensor & Scalar & Nonparametric & MC\\
 		    Yang and Dunson~\cite{Yang2016} & Vector (Categorical) & Scalar (Categorical) & Tucker & Gibbs\\
 		    CATCH~\cite{Pan2018P} & Tensor+Vector & Scalar (Categorical) & Tucker & MLE\\\hline
 		    BTRR~\cite{Guhaniyogi2021} & Vector & Tensor & CP & Gibbs\\
 		    Spencer et al.~\cite{Spencer2019, Spencer2020} & Vector & Tensor & CP & Gibbs\\
 		    SGTM~\cite{Guha2021} & Vector & Symmetric Tensor & CP & Gibbs\\
                BSTN~\cite{Lee2022} & Vector & Tensor & Other & Gibbs\\
 		    SGPRN~\cite{Li2020} & Matrix & Tensor & Nonparametric & VI\\
 		    MLTR~\cite{Hoff2015} & Tensor & Tensor & Tucker & Gibbs\\
 		    ART~\cite{Billio2022} & Tensor & Tensor & CP & Gibbs\\\hline
        \end{tabular}
 	    Gibbs: MCMC with Gibbs sampling. MAP: Maximum a posteriori.
 	    MC: Monte Carlo Method. MLE: Maximum likelihood estimator. 
            OLGP: Online local Gaussian process~\cite{OLGP1, OLGP2}. VI: Variational Inference.
    \end{minipage}
 \end{table*}

\subsection{Bayesian Tensor Predictor Regression}\label{BTPR}
In recent years, Bayesian tensor predictor regression models have gained an increasing attention. For example, Suzuki~\cite{Suzuki2015} develop a Bayesian framework based on the basic tensor linear regression model
\begin{equation} Y_i=\langle \mathcal{W}, \mathcal{X}_i\rangle+\epsilon_i, \label{Suzuki}\end{equation}
where $Y_i\in\mathbb{R}$ is a univariate response, $\mathcal{X}_i\in\mathbb{R}^{M_1\times\cdots\times M_K}$ is a tensor-valued predictor, $\mathcal{W}\in\mathbb{R}^{M_1\times\cdots\times M_K}$ is the coefficient tensor, and $\langle\cdot,\cdot\rangle$ is the tensor inner product (\ref{inner product}). The error terms $\epsilon_i$'s are assumed i.i.d. following a normal distribution $\mathcal{N}(0, \sigma^2)$. To achieve parsimony in free parameters, a rank-$r$ CP structure (\ref{CP}) is imposed on the coefficient tensor $\mathcal{W}$: 
$$\mathcal{W}=[\![\boldsymbol{U}^{(1)},...,\boldsymbol{U}^{(K)}]\!],$$
where $\boldsymbol{U}^{(k)}\in\mathbb{R}^{r\times M_K}$ ($k=1,2,...,K$) are latent factors. To complete model specification, a Gaussian prior is placed on the latent matrices:
$$ \pi(\boldsymbol{U}^{(1)},...,\boldsymbol{U}^{(K)}|r)\propto \exp\Big\{-\frac{r}{2\sigma^2_p}\sum_{k=1}^K \text{Tr}[\boldsymbol{U}^{(k)^{\top}}\boldsymbol{U}^{(k)}]\Big\},$$
and an independent prior is used for the rank $r$:
$$ \pi(r)=\frac{1}{N_{\xi}}\xi^{r(M_1+\cdots+M_K)},$$
where $0<\xi<1$ is a positive real number, and $N_{\xi}$ is the normalizing constant.

In order to adjust for other covariates in the model and accommodate various data types of the response variable, Guhaniyogi et al.~\cite{Guhaniyogi2017} propose a Bayesian method based on the generalized tensor predictor regression model~\eqref{eq2}. Given a scalar response $y$, vectorized predictors $\boldsymbol{z}\in\mathbb{R}^p$ and a tensor predictor $\mathcal{X}\in \mathbb{R}^{p_1\times p_2\times...\times p_D}$, the regression model is given by
\begin{equation} y\sim f(\alpha+\boldsymbol{z}^{\top}\boldsymbol{\gamma}+\langle \mathcal{X}, \mathcal{B}\rangle, \sigma), \label{Guhaniyogi1}\end{equation}
where $f(\mu, \sigma)$ is a family of distributions with location $\mu$ and scale $\sigma$, $\boldsymbol{\gamma}\in\mathbb{R}^{p}$ are coefficients for predictors $\boldsymbol{z}$, $\mathcal{B}\in\mathbb{R}^{p_1\times p_2\times...\times p_D}$ is the coefficient tensor, and $\langle\cdot,\cdot\rangle$ is the tensor inner product (\ref{inner product}). A CP structure is imposed on the tensor coefficient $\mathcal{B}$:
$$ \mathcal{B}=\sum_{r=1}^R \boldsymbol{\beta}_1^{(r)}\circ\cdots\circ\boldsymbol{\beta}_D^{(r)}.$$

Under the Bayesian framework, Guhaniyogi et al.~\cite{Guhaniyogi2017} propose a multiway Dirichlet generalized double Pareto (M-DGDP) prior over the latent factors $\boldsymbol{\beta}_j^{(r)}$. This prior promotes the joint shrinkage on the global and local component parameters, as well as accommodates dimension reduction by favoring low-rank decompositions. Specifically, the M-DGDP prior first assigns a multivariate Gaussian prior on $\boldsymbol{\beta}_j^{(r)}$:
\begin{equation} 
\boldsymbol{\beta}_j^{(r)}\sim \mathcal{N}(\boldsymbol{0},(\phi_r\tau)\boldsymbol{W}_{jr}), ~j=1,\ldots,D.
\label{eqn2}\end{equation}
The shrinkage across components is induced in an exchangeable way, with a global scale parameter $\tau\sim\text{Gamma}(a_{\tau}, b_{\tau})$ adjusted in each component by $\phi_r$ for $r=1,2,...,R$, where $\Phi=(\phi_1,...,\phi_R)\sim \text{Dirichlet}(\alpha_1,...,\alpha_R)$ encourages shrinkage towards lower ranks in the CP structure. In addition, $\boldsymbol{W}_{jr}=\text{diag}(w_{jr,1},\cdots,w_{jr,p_j})$, $j=1,2,...,D$ and $r=1,2,...,R$, are scale parameters for each component, where a hierarchical prior is used,
\begin{equation}
    w_{jr, k}\sim\text{Exp}(\lambda^2_{jr}/2),\quad \lambda_{jr}\sim\text{Gamma}(a_{\lambda},b_{\lambda}).
    \label{eq22}
\end{equation}

In the M-DGDP prior, flexibility in estimating $\mathcal{B}_r=\{\boldsymbol{\beta}_j^{(r)};1\leq j\leq D\}$ is achieved by modeling individual-level heterogeneity via element-specific scaling parameters $w_{jr,k}$'s. The common rate parameter $\lambda_{jr}$ shares information between individual elements, hence leads to shrinkage at the local scale.

Besides linear models, a class of Gaussian process (GP) based nonparametric approaches have been proposed to model nonlinear relationships in the tensor-valued predictors. Given a dataset of $N$  paired observations $\mathcal{D}=\{(\mathcal{X}_n, y_n)|n=1,2,...,N\}$, Zhao et al.~\cite{Zhao2014P} aggregate all $N$ tensor inputs $\mathcal{X}_n~(n=1,2,...,N)$ into a design tensor $\mathcal{X}\in\mathbb{R}^{N\times I_1\times\cdots\times I_M}$, and collect the responses in the vector form $\boldsymbol{y}=[y_1,...,y_N]^{\top}$. The distribution of the response vector can be factored over the observations as
\begin{equation} 
\boldsymbol{y}\sim \prod_{n=1}^N \mathcal{N}(y_n|f(\mathcal{X}_n),\sigma^2).
\label{eq24}
\end{equation}
Here $f(\cdot)$ is a latent function on which a GP prior is placed 
\begin{equation} 
f(\mathcal{X})\sim \text{GP}(m(\mathcal{X}), k(\mathcal{X}, \mathcal{X}')|\boldsymbol{\theta}),
\label{eq25}
\end{equation}
where $k(\mathcal{X},\mathcal{X}')$ is the covariance function (kernel), $\boldsymbol{\theta}$ is the associated hyperparameter vector, and $m(\mathcal{X})$ is the mean function which is set to be zero in~\cite{Zhao2014P}. The authors further propose to use the following product kernel in \eqref{eq25}:
\begin{equation}
    k(\mathcal{X}, \mathcal{X}')=\alpha^2\prod_{d=1}^D \exp(\frac{D(p(\boldsymbol{x}|\Omega_d^{\mathcal{X}})~\Vert~q(\boldsymbol{x}'|\Omega_d^{\mathcal{X}'}))}{-2\beta_d^2}),
    \label{eq23}
\end{equation}
where $\alpha$ is a magnitude hyperparameter, $\beta_d$ denotes the $d$-mode length-scale hyper-parameter, and $D$ is the symmetric Kullback-Leibler (KL) divergence defined as
$$D(P||Q)=\text{KL}(P||Q)+\text{KL}(Q||P).$$ 
The distributions $p$ and $q$ in the symmetric KL divergence are characterized by the hyper-parameters $\Omega_d$, which can be estimated from the $d$-mode unfolding matrix $\boldsymbol{X}_d$ of tensor $\mathcal{X}$ by treating each $\boldsymbol{X}_d$ as a generative model with $I_d$ variables and $I_1\times\cdots\times I_{d-1}\times I_{d+1}\times\cdots\times I_D$ observations. Given the prior construction, the hyperparameters $\boldsymbol{\theta}=\{\alpha,\beta_d|d=1,2,...,D\}$ and $\sigma$ are then estimated by maximum a posteriori (MAP). While the computational complexity of GP-based methods is usually excessive, Hou et al.~\cite{Hou2015} take advantage of the online local Gaussian Process (OLGP) and present a computationally-efficient approach for the nonparametric model in \eqref{eq24}-\eqref{eq23}.

To further mitigate the burden of high-dimensionality, Imaizumi and Hayashi~\cite{Imaizumi2016} propose an additive-multiplicative nonparametric regression (AMNR) method that concurrently decomposes the functional space and the input space. This method is referred to as a doubly decomposing nonparametric tensor regression method.

Denote a Sobolev space by $\mathcal{W}^{\beta}(\mathcal{X})$, which is a space of $\beta$-times differentiable functions with the support $\mathcal{X}$. Let $\mathcal{X}=\bigotimes_k \boldsymbol{x}_k:=\boldsymbol{x}_1\otimes\cdots\otimes \boldsymbol{x}_K$ be a rank-one tensor denoted by the outer product of vectors $\boldsymbol{x}_k\in\mathcal{X}^{(k)}$ ($\otimes$ is the outer product). Let $f\in\mathcal{W}^{\beta}(\bigotimes_k\mathcal{X}^{(k)})$ be a function on a rank-one tensor. For any $f$ we can construct $\tilde{f}(\boldsymbol{x}_1,...,\boldsymbol{x}_K)\in\mathcal{W}^{\beta}(\mathcal{X}^{(1)}\times\cdots\times\mathcal{X}^{(k)})$ such that $\tilde{f}(\boldsymbol{x}_1,...,\boldsymbol{x}_K)=f(\mathcal{X})$ using function decomposition as $\tilde{f}=f\circ h$ with $h:(\boldsymbol{x}_1,...,\boldsymbol{x}_K)\to \bigotimes_k \boldsymbol{x}_k$. Then $f$ can be decomposed into a set of local functions $\{f_m^{k}\in\mathcal{W}^{\beta}(\mathcal{X}^{(k)})\}_m$ following \cite{Hackbusch}:
\begin{equation} 
f(\mathcal{X})=\tilde{f}(\boldsymbol{x}_1,...,\boldsymbol{x}_K)=\sum_{m=1}^{M}\prod_{k=1}^K f_m^{(k)}(\boldsymbol{x}_{k}),
\label{eq80}
\end{equation}
where $M$ represents the complexity of $f$ (i.e., the ``rank'' of the model).

Based on (\ref{eq80}), for a rank-$R$ tensor $\mathcal{X}$, Imaizumi and Hayashi~\cite{Imaizumi2016} define the AMNR function as:
\begin{equation} f^{AMNR}(\mathcal{X}):=\sum_{m=1}^M \sum_{r=1}^R \lambda_r \prod_{k=1}^K f_m^{(k)}(\boldsymbol{x}_r^{(k)}),\label{AMNR}\end{equation}
which is obtained by first writing a rank-$R$ tensor as the sum of $R$ rank-one tensors, and then decomposing the function into a set of local functions for each rank-one tensor. Under the Bayesian framework, a GP prior is assigned to the local functions $f_m^{(k)}$, and the Gaussian distribution (\ref{eq24}) is utilized to associate the scalar response $Y_i$ with the function $f^{AMNR}(\mathcal{X}_i)$.

While the previous studies mainly deal with  regression problems with continuous response variables, the probabilistic methods can also apply to categorical-response regression problems with tensor-valued predictors, i.e., the tensor classification problems. For example, Pan et al.~\cite{Pan2018P} propose a covariate-adjusted tensor classification model (CATCH), which jointly models the relationship among the covariates, tensor predictors, and categorical responses. Given a categorical response $Y\in\{1,2,...,K\}$, a vector of covariates $\boldsymbol{U}\in\mathbb{R}^q$, and tensor-variate predictors $\mathcal{X}\in\mathbb{R}^{p_1\times\cdots\times p_M}$, the CATCH model is proposed as
\begin{gather}
    \boldsymbol{U}|(Y=k)\sim \mathcal{N}(\boldsymbol{\Phi}_k,\boldsymbol{\Psi}) \label{CATCH1}\\
    \mathcal{X}|(\boldsymbol{U}=\boldsymbol{u},Y=k)\sim \text{TN}(\boldsymbol{\mu}_k+\boldsymbol{\alpha}\bar{\times}_{(M+1)}\boldsymbol{u};\boldsymbol{\Sigma}_1,...,\boldsymbol{\Sigma}_M), \label{CATCH2}
\end{gather}
where $\boldsymbol{\Phi}_k\in\mathbb{R}^q, \boldsymbol{\Psi}\in\mathbb{R}^{q\times q}$  is positive definite, $\boldsymbol{\alpha}\in\mathbb{R}^{p_1\times...\times p_M\times q}, \boldsymbol{\mu}_k\in\mathbb{R}^{p_1\times...\times p_M}$, and $\boldsymbol{\Sigma}_m\in\mathbb{R}^{p_m\times p_m}$ is positive definite for $m=1,...,M$. Here TN$(\cdot)$ is the tensor normal distribution, and $\bar{\times}_{(M+1)}$ is the $(M+1)$-mode tensor vector product. 

In equation (\ref{CATCH1}), it is assumed that $\{Y,\boldsymbol{U}\}$ follow a classical LDA model, where $\boldsymbol{\Phi}_k$ is the mean of $\boldsymbol{U}$ within class $k$ and $\boldsymbol{\Psi}$ is the common within class covariance of $\boldsymbol{U}$. Similarly, in equation (\ref{CATCH2}) a common within class covariance structure of $\mathcal{X}$ is assumed (denoted by $\boldsymbol{\Sigma}_m, m=1,2,...,M$), which does not depend on $Y$ after adjusting for the covariates $\boldsymbol{U}$. The tensor coefficient $\boldsymbol{\alpha}$ characterizes the linear dependence of tensor predictor $\mathcal{X}$ on the covariates $\boldsymbol{U}$, and $\boldsymbol{\mu}_k$ is the covariate-adjusted within-class mean of $\mathcal{X}$ in class $k$.

While the goal is to predict $Y$ given $\{\boldsymbol{U}, \mathcal{X}\}$, based on the Bayes' rule the optimal classifier under the CATCH model is derived by maximizing the posterior probability
\begin{equation}
\hat{Y}=\arg\max_{k=1,2,...,K}P(Y=k|\mathcal{X}=\boldsymbol{x},\boldsymbol{U}=\boldsymbol{u})
=\arg\max_{k=1,2,...,K}\pi_kf_k(\boldsymbol{x},\boldsymbol{u}),
\label{CATCH3}\end{equation}
where $\pi_k=P(Y=k)$ and $f_k(\boldsymbol{x}, \boldsymbol{u})$ is the joint density function of $\mathcal{X}$ and $\boldsymbol{U}$ conditional on $Y=k$. Combining \eqref{CATCH1} and \eqref{CATCH2}, equation (\ref{CATCH3}) is transformed into
$$ \hat{Y}=\arg\max_{k=1,2,...,K}\{a_k+\boldsymbol{\gamma}_k^{\top}\boldsymbol{U}+\langle\mathcal{B}_k,\mathcal{X}-\boldsymbol{\alpha}\bar{\times}_{(M+1)}\boldsymbol{U}\rangle\},$$
where $\boldsymbol{\gamma}_k=\boldsymbol{\Psi}^{-1}(\boldsymbol{\Phi}_k-\boldsymbol{\Phi}_1), \mathcal{B}_k=[\![\boldsymbol{\mu}_k-\boldsymbol{\mu}_1;\boldsymbol{\Sigma}_1^{-1},...,\boldsymbol{\Sigma}_M^{-1}]\!]$ following a Tucker structure with the core tensor $\boldsymbol{\mu}_k-\boldsymbol{\mu}_1$ and latent matrices $\boldsymbol{\Sigma}_1^{-1},...,\boldsymbol{\Sigma}_M^{-1}$, and $a_k=\log(\pi_k/\pi_1)-\frac{1}{2}\boldsymbol{\gamma}_k^{\top}(\boldsymbol{\Phi}_k+\boldsymbol{\Phi}_1)-\langle \mathcal{B}_k,\frac{1}{2}(\boldsymbol{\mu}_k+\boldsymbol{\mu}_1)\rangle$ is a scalar that does not depend on $\mathcal{X}$ or $\boldsymbol{U}$.

Given i.i.d. samples $\{Y^i,\boldsymbol{U}^i, \mathcal{X}^i\}_{i=1}^n$, the parameters $\{\pi_k,\boldsymbol{\Phi}_k,\boldsymbol{\gamma}_k,\boldsymbol{\mu}_k, \mathcal{B}_k\}_{k=1}^K$ and $\{\boldsymbol{\Sigma}_m\}_{m=1}^M$ can be estimated to build an accurate classifier based on the data. Regularization is used when estimating $\mathcal{B}_k$ in order to facilitate sparsity.

Though not modeling tensor predictors, Yang and Dunson~\cite{Yang2016} employ tensor methods to deal with classification problems with categorical predictors. Specifically, ~\cite{Yang2016} develop a framework for nonparametric Bayesian classification through performing decomposition on the tensor constructed from the conditional probability 
$$P(Y=y|X_1=x_1,...,X_p=x_p),$$
with a categorical response $Y\in\{1,2,...,d_0\}$ and a vector of $p$ categorical predictors $\boldsymbol{X}=(X_1,X_2,...,X_p)^{\top}$. The conditional probability can be structured as a $d_0\times d_1\times\cdots\times d_p$-dimensional tensor, where $d_j$ $(j=1,2,...,p)$ denotes the number of levels of the $j$th categorical predictor $X_j$. This tensor is called a conditional probability tensor, and the set of all conditional probability tensors is denoted by $\mathcal{P}_{d_1,...,d_p}(d_0)$. Therefore, $\mathcal{P}\in\mathcal{P}_{d_1,...,d_p}(d_0)$ implies
\begin{gather*} 
    \mathcal{P}_{y,x_1,...,x_p}\geq 0\quad\text{for every}~ y,x_1,...,x_p;\\
    \sum_{y=1}^{d_0}\mathcal{P}_{y,x_1,...,x_p}=1\quad\text{for every}~ x_1,...,x_p.
\end{gather*}

Since all the conditional probabilities are entries in the conditional probability tensor, the classification problem is converted into a tensor decomposition problem. Additionally, Yang and Dunson~\cite{Yang2016} prove that every conditional probability tensor $\mathcal{P}\in\mathcal{P}_{d_1,...,d_p}(d_0)$ can be expressed by a Tucker structure 
\begin{equation*}
\mathcal{P}_{y,x_1,...,x_p}=P(y|x_1,...,x_p)
=\sum_{h_1=1}^{k_1}\cdots\sum_{h_p=1}^{k_p}\lambda_{h_1h_2...h_p}(y)\prod_{j=1}^p \pi_{h_j}^{(j)}(x_j),
\end{equation*}
with all positive parameters satisfying
\begin{equation*}\begin{split}
& \sum_{c=1}^{d_0}\lambda_{h_1h_2...h_p}(c)=1, \quad\text{for every}~ h_1,h_2,...,h_p,\\
& \sum_{h=1}^{k_j} \pi_h^{(j)}(x_j)=1, \quad\text{for every pair of }~ j,x_j.
\end{split}\end{equation*}

The inference of the Tucker coefficients is carried out under the Bayesian framework. Specifically, independent Dirichlet priors are assigned to the parameters $\boldsymbol{\Lambda}=\{\lambda_{h_1,...,h_p}(c), c=1,2,...,d_0\}$ and $\boldsymbol{\pi}=\{\pi_{h_j}^{(j)}(x_j),h_j=1,2,...,k_j\}$ ($x_j=1,2,...,d_j, h_j=1,2,...,k_j, j=1,2,...,p$):
\begin{align*}
  &  \bigg\{ \lambda_{h_1,...,h_p}(1),...,\lambda_{h_1,...,h_p}(d_0)\bigg\}\sim \text{Dirichlet}(\frac{1}{d_0},...,\frac{1}{d_0}),\\
  &  \bigg\{\pi_1^{(j)}(x_j),...,\pi_{k_j}^{(j)}(x_j)\bigg\}\sim \text{Dirichlet}(\frac{1}{k_j},...,\frac{1}{k_j}),~j=1,...,p. 
\end{align*}
These priors impose the non-negativity and sum-to-one constraints naturally and lead to conditional conjugacy in posterior computation. Additionally, \cite{Yang2016} assign priors on the hyper-parameters in the Dirichlet priors to promote a fully Bayesian treatment. These priors place most of the probability on few elements to induce sparsity in their vectors.

\subsection{Bayesian Tensor Response Regression}\label{BTRR}
Guhaniyogi and Spencer~\cite{Guhaniyogi2021} propose a Bayesian regression model with a tensor response and scalar predictors. Let $\mathcal{Y}_t\in\mathbb{R}^{p_1\times p_2\times...\times p_D}$ be a tensor-valued response, and $\boldsymbol{x}_t=(x_{1,t},...,x_{m,t})\in\mathcal{X}\subset\mathbb{R}^m$ be an $m$-dimensional vector predictor measured at time $t$. Assuming that both the response $\mathcal{Y}_t$ and the predictors $\boldsymbol{x}_t$ are centered around their respective means, the proposed regression model for $\mathcal{Y}_t$ on $\boldsymbol{x}_t$ is given by
\begin{equation} 
\mathcal{Y}_t=\boldsymbol{\Gamma}_1 x_{1,t}+\cdots+\boldsymbol{\Gamma}_m x_{m,t}+\mathcal{E}_t,\quad i=1,2,...,n,
\label{eq19}
\end{equation}
where $\boldsymbol{\Gamma}_k\in\mathbb{R}^{p_1\times p_2\times...\times p_D},k=1,2,...,m$ is the tensor coefficient corresponding to the predictor $x_{k,t}$, and $\mathcal{E}_t\in\mathbb{R}^{p_1\times p_2\times...\times p_D}$ represents the error tensor. To account for the temporal correlation in the response tensor, the error tensor $\mathcal{E}_t$ is assumed to follow a component-wise AR(1) structure across $t$: vec$(\mathcal{E}_t)=\kappa\text{vec}(\mathcal{E}_{t-1})+\text{vec}(\boldsymbol{\eta}_t)$, where $\kappa\in (-1,1)$ is the correlation coefficient, and $\boldsymbol{\eta}_t\in\mathbb{R}^{p_1\times p_2\times...\times p_D}$ is a random tensor, with each entry following a Gaussian distribution $\mathcal{N}(0, \sigma^2/(1-\kappa^2))$.

Next, a CP structure is imposed on each $\boldsymbol{\Gamma}_k$ to reduce the dimensionality of coefficient tensors, i.e., $\boldsymbol{\Gamma}_k=\sum_{r=1}^R \boldsymbol{\gamma}_{1,k}^{(r)}\circ\cdots\circ\boldsymbol{\gamma}_{D,k}^{(r)}$. Although Guhaniyogi et al's previously proposed M-DGDP prior~(\ref{eqn2})(\ref{eq22}) over the latent factors $\boldsymbol{\gamma}_{j,k}^{(r)}$ can promote global and local sparsity, Guhaniyogi and Spencer~\cite{Guhaniyogi2021} claim that a direct application of M-DGDP prior leads to inaccurate estimation due to a less desirable tail behavior of the coefficient distributions. Instead, a multiway stick breaking shrinkage prior (M-SB) is assigned to $\boldsymbol{\gamma}_{j,k}^{(r)}$, where the main difference compared to the M-DGDP prior is how shrinkage is achieved across ranks. The construction of the M-SB prior is given as follows. Let $\boldsymbol{W}_{jr,k}=\text{diag}(w_{jr,k,1},...,w_{jr,k,p_d})$. Then we set
$$ \boldsymbol{\gamma}_{j,k}^{(r)}\sim\mathcal{N}(0, \tau_{r,k}\boldsymbol{W}_{jr,k}).$$
Further set $\tau_{r,k}=\phi_{r,k}\tau_k$ to be scaling specific to rank $r$ ($r=1,...,R$). Then effective shrinkage across ranks is achieved by adopting a stick breaking construction for the rank-specific parameter $\phi_{r,k}$:
\begin{align*}
& \phi_{r,k}=\xi_{r,k}\prod_{l=1}^{r-1}(1-\xi_{l,k}),\quad r=1,...,R-1,\\
& \phi_{R,k}=\prod_{l=1}^{R-1}(1-\xi_{l,k}),
\end{align*}
where $\xi_{r,k}\sim_{iid}\text{Beta}(1,\alpha_k).$ The Bayesian setting is then completed by specifying
\begin{equation*} 
\tau_k\sim\text{InvGamma}(a_{\tau}, b_{\tau}),~~w_{jr,k,i}\sim\text{Exp}(\lambda_{jr,k}^2/2),
~~\lambda_{jr,k}\sim\text{Gamma}(a_{\lambda},b_{\lambda}),
\end{equation*}
where the hierarchical prior of $w_{jr,k,i}$ allows the local scale parameters $\boldsymbol{W}_{jr,k}$ to achieve individual-level shrinkage.

Based on the regression function (\ref{eq19}), Spencer et al.~\cite{Spencer2019, Spencer2020} consider a brain imaging application and develop an additive mixed effect model that simultaneously measures the activation due to stimulus at voxels in the $g$th brain region and connectivity among $G$ brain regions. Let $\mathcal{Y}_{i,g,t}\in\mathbb{R}^{p_{1,g}\times\cdots\times p_{D,g}}$ be the tensor of observed fMRI data in brain region $g$ for the $i$th subject at the $t$th time point, and $x_{1,i,t},...,x_{m,i,t}\in\mathbb{R}$ be the activation-related predictors. The regression function is given by
$$ \mathcal{Y}_{i,g,t}=\boldsymbol{\Gamma}_{1,g}x_{1,i,t}+\cdots\boldsymbol{\Gamma}_{m,g}x_{m,i,t}+d_{i,g}+\mathcal{E}_{i,g,t}$$
for subject $i=1,2,...,n$ in region $g=1,2,...,G$ and time $t=1,2,...,T$. Here $\mathcal{E}_{i,g,t}\in\mathbb{R}^{p_{1,g}\times\cdots\times p_{D,g}}$ is the error tensor, of which the elements are assumed to follow a normal distribution with zero mean and shared variance $\sigma_y^2$. $\boldsymbol{\Gamma}_{k,g}\in\mathbb{R}^{p_{1,g}\times\cdots\times p_{D,g}}$ represents activation due to the $k$th stimulus at $g$th brain region. Each $\boldsymbol{\Gamma}_{k,g}$ is assumed to follow a CP structure, and an M-SB prior is assigned to the latent factors of the CP decomposition to determine the nature of activation. Also, $d_{i,g}\in\mathbb{R}$ are region- and subject-specific random effects that are jointly modeled to borrow information across regions of interest. Specifically, a Gaussian graphical LASSO prior is imposed on these random effects:
\begin{gather*} 
\boldsymbol{d}_i=(d_{i,1},...,d_{i,G})^{\top}\sim\mathcal{N}(\boldsymbol{0},\boldsymbol{\Omega}^{-1}),\quad i=1,2,...,n,\\
p(\boldsymbol{\omega}|\zeta)=C^{-1}\prod_{g<g_1}[DE(\omega_{gg_1}|\zeta)]\prod_{g=1}^G[\text{Exp}(\omega_{gg}|\frac{\zeta}{2})]\boldsymbol{1}_{\boldsymbol{\Omega}\in \mathcal{P}^{+}},
\end{gather*}
where $\mathcal{P}^+$ is the class of all positive definite matrices and $C$ is a normalization constant. The covariance $\boldsymbol{\omega}=(\omega_{gg_1}:g\leq g_1)$ is a vector of upper triangle and diagonal entries of the precision matrix $\boldsymbol{\Omega}$. By properties of the multivariate Gaussian distribution, a small value of $\omega_{gg_1}$ stands for weak connectivity between regions of interest (ROIs) $g$ and $g_1$, given other ROIs. In practice, a double exponential prior is employed on the off-diagonal entries of the precision matrix $\boldsymbol{\Omega}$ to favor shrinkage among these entries. A full Bayesian prior construction is completed by assigning a Gamma prior on $\zeta$ and an inverse Gamma prior on the variance parameter $\sigma_y^2$.

To study brain connectome datasets acquired using diffusion weighted magnetic resonance imaging (DWI), Guha and Guhaniyogi~\cite{Guha2021} propose a generalized Bayesian linear model with a symmetric tensor response and scalar predictors. Let $\mathcal{Y}_i\in\mathcal{Y}\subset\mathbb{R}^{p\times...\times p}$ be a symmetric tensor response with diagonal entries being zero, $\boldsymbol{x}_i=(x_{i1},...,x_{im})^{\top}$ be $m$ predictors of interest, and $\boldsymbol{z}_i=(z_{i1},...,z_{il})^{\top}$ be $l$ auxiliary predictors corresponding to the $i$th individual. Let $\mathcal{J}=\{\boldsymbol{j}=(j_1,...,j_D):1\leq j_1<\cdots<j_D\leq p\}$ be a set of indices. Given that $\mathcal{Y}_i$ is symmetric with dummy diagonal entries, it suffices to build a probabilistic generative mechanism for $y_{i,\boldsymbol{j}}~(\boldsymbol{j}\in\mathcal{J})$. In practice, a set of conditionally independent generalized linear models are utilized. Let $E(y_{i,\boldsymbol{j}})=\omega_{i,\boldsymbol{j}}$, for $\boldsymbol{j}\in\mathcal{J}$, we have
\begin{equation*}
\omega_{i,\boldsymbol{j}}=
H^{-1}(\beta_0+B_{1,\boldsymbol{j}}x_{i1}+\cdots+B_{m,\boldsymbol{j}}x_{im}+\beta_1 z_{i1}+\cdots+\beta_lz_{il}),
\end{equation*}
where $B_{1,\boldsymbol{j}},...,B_{m,\boldsymbol{j}}$ respectively represents the entry $\boldsymbol{j}=(j_1,...,j_D)$ of the $p\times\cdots\times p$ symmetric coefficient tensors $\mathcal{B}_{1},...,\mathcal{B}_m$ with diagonal entries zero, $\beta_0,\beta_1,...,\beta_l\in\mathbb{R}$ are the intercept and coefficients corresponding to variables $z_{i1},...,z_{il}$, respectively, and $H(\cdot)$ is the link function. The model formulation implies a similar effect of any of the auxiliary variables $(z_{i1},...,z_{il})$ on all entries of the response tensor but varying effects of the $h$th predictor on different entries $\boldsymbol{j}\in\mathcal{J}$ of the response tensor. To account for associations between tensor nodes and predictors and to achieve parsimony in tensor coefficients, a CP-like structure is imposed on symmetric coefficient tensors $\mathcal{B}_1,...,\mathcal{B}_m$, i.e.,
\begin{equation}
    B_{h,\boldsymbol{j}}=\sum_{r=1}^R \lambda_{h,r}u_{h,j_1}^{(r)}\cdots u_{h,j_D}^{(r)},\quad h=1,2,...,m;~\boldsymbol{j}\in\mathcal{J},
    \label{eq20}
\end{equation}
where $\boldsymbol{u}_h^{(r)}=(u_{h,1}^{(r)},...,u_{h,p}^{(r)})^{\top}\in\mathbb{R}^p$ are latent factors and $\lambda_{h,r}\in\{0,1\}$ is a binary inclusion variable determining if the $r$th summand in (\ref{eq20}) is relevant in model setting. Further let $\tilde{\boldsymbol{u}}_{h,k}=(u_{h,k}^{(1)},...,u_{h,k}^{(R)})$, then the $h$th predictor of interest is considered to have no impact on the $k$th tensor if $\tilde{\boldsymbol{u}}_{h,k}=0$. In order to directly study the effect of tensor nodes related to the $h$th predictor of interest, a spike-and-slab mixture distribution prior is assigned on $\tilde{\boldsymbol{u}}_{h,k}$:
\begin{equation*} 
\tilde{\boldsymbol{u}}_{h,k}\sim \begin{cases} \mathcal{N}(\boldsymbol{0},\boldsymbol{M}_h), & \text{if } \eta_{h,k}=1 \\\delta_{\boldsymbol{0}}, & \text{if } \eta_{h,k}=0 \end{cases},~~ \eta_{h,k}\sim\text{Bern}(\xi_h),~~\boldsymbol{M}_h\sim IW(\boldsymbol{S},\nu),~~\xi_h\sim U(0,1),
\end{equation*}
where $\delta_{\boldsymbol{0}}$ is the Dirac function at $\boldsymbol{0}$ and $\boldsymbol{M}_h$ is a covariance matrix of order $R\times R$. Here $IW(\boldsymbol{S}, \nu)$ denotes an Inverse-Wishart distribution with an $R\times R$ positive definite scale matrix $\boldsymbol{S}$ and $\nu$ degrees of freedom. The parameter $\xi_h$ corresponds to the probability of the nonzero mixture component and $\eta_{h,k}$ is a binary indicator that equals $0$ if $\tilde{\boldsymbol{u}}_{h,k}=\delta_{\boldsymbol{0}}$. Thus, the posterior distributions of $\eta_{h,k}$'s can help identify nodes related to a chosen predictor.

To impart increasing shrinkage on $\lambda_{h,r}$ as $r$ grows, a hierarchical prior is imposed on $\lambda_{h,r}$: 
$$\lambda_{h,r}\sim\text{Bern}(\nu_{h,r}),~\nu_{h,r}\sim \text{Beta}(1, r^{\zeta}), \zeta>1.$$
In addition, a Gaussian prior $\mathcal{N}(a_{\beta}, b_{\beta})$ is placed on $\beta_0,\beta_1,...,\beta_l$.

Recently, Lee et al.~\cite{Lee2022} develop a Bayesian skewed tensor normal (BSTN) regression, which addresses the problem of considerable skewness in the tensor response in a study of periodontal disease (PD). For an order-$K$ tensor response $\mathcal{Y}_i\in\mathbb{R}^{d_1\times\cdots\times d_K}$ with a vector of covariates $\boldsymbol{x}_i\in\mathbb{R}^p$, the regression model is given by
$$\mathcal{Y}_i=\mathcal{B}\bar{\times}_{(K+1)}\boldsymbol{x}_i+\mathcal{E}_i,\quad\text{for }i=1,2,...,n,$$
where $\mathcal{B}\in\mathbb{R}^{d_1\times\cdots\times d_K\times p}$ is an order-$(K+1)$ coefficient tensor, $\bar{\times}_{(K+1)}$ is the $(K+1)$th mode vector product, and $\mathcal{E}_i\in\mathbb{R}^{d_1\times\cdots\times d_K}$ is the error tensor. The skewness in the distribution of $\mathcal{Y}$ is modeled by 
$$ \mathcal{E}_i=|\mathcal{Z}_{2i}|\times_K\boldsymbol{\Lambda}+\mathcal{Z}_{1i},$$
where $\boldsymbol{\Lambda}=\text{diag}(\lambda_1,...,\lambda_{d_K})\in\mathbb{R}^{d_K\times d_K}$ is a digonal matrix with skewness parameters $\boldsymbol{\lambda}=(\lambda_1,...,\lambda_{d_K})$, $|\boldsymbol{M}|$ denotes a matrix whose elements are absolute values of the corresponding elements in matrix $\boldsymbol{M}$, and $\times_K$ is the mode-$K$ tensor matrix product. The tensor  $\mathcal{Z}_{2i}\in\mathbb{R}^{d_1\times\cdots\times d_K}$ follows a tensor normal distribution $\mathcal{Z}_{2i}\sim \text{TN}(\boldsymbol{0};\boldsymbol{I}_{d_1},...,\boldsymbol{I}_{d_{K-1}},\boldsymbol{D}_{\boldsymbol{\sigma}}^2)$, and is assumed to be independent of $\mathcal{Z}_{1i}\sim\text{TN}(\boldsymbol{0};\boldsymbol{R}_1,...,\boldsymbol{R}_{K-1}, \boldsymbol{D}_{\boldsymbol{\sigma}}\boldsymbol{R}_K\boldsymbol{D}_{\boldsymbol{\sigma}})$, where $\boldsymbol{R}_1,...,\boldsymbol{R}_K$ are positive-definite correlation matrices, and $\boldsymbol{D}_{\boldsymbol{\sigma}}=\text{diag}(\sigma_1,...,\sigma_{d_K})$ is a diagonal matrix of positive scale parameters $\sigma_1,...,\sigma_{d_K}$. The parameterization for the tensor normal $\mathcal{Z}_{1i}$ via correlation matrices $\boldsymbol{R}_1,...,\boldsymbol{R}_K$ avoids the common identifiability issue. Only the $K$th mode of $\mathcal{Z}_{2i}$ is multiplied by a skewness matrix $\boldsymbol{\Lambda}=\text{diag}(\lambda_1,...,\lambda_{d_K})$ because the skewness level is assumed to be the same in all combinations of the first $(K-1)$ modes in the PD dataset. When $\lambda_j$ is positive (or negative), the corresponding marginal density of $y_{i_1,...,i_{K-1},j}$ of tensor response $\mathcal{Y}$ is skewed to the right (left).

Various prior distributions can be put on the parameters. For example, an independent zero-mean normal density with a pre-specified variance is utilized as the common prior for $\boldsymbol{\lambda}=(\lambda_1,...,\lambda_{d_K})$, and common independent inverse-gamma distributions $IG(g_1,g_2)$ with pre-specified shape $g_1>0$ and scale $g_2>0$ are imposed on $\boldsymbol{\sigma}=(\sigma_1,...,\sigma_{d_K})$. The parametric correlation matrices $\boldsymbol{R}_1,...,\boldsymbol{R}_{K}$ are assumed to be equicorrelation matrices with independent uniform priors $Unif(-1,1)$ for unknown off-diagonal elements. A tensor normal distribution $\text{TN}(\boldsymbol{0}; \boldsymbol{C}_1,...,\boldsymbol{C}_{K+1})$ with zero mean and known covariance matrices $\boldsymbol{C}_1,...,\boldsymbol{C}_{K+1}$ is put on the tensor coefficient $\mathcal{B}$. Lee et al.~\cite{Lee2022} also propose an alternative prior distribution for $\mathcal{B}$, where a spike-and-slab prior is employed to introduce sparsity.

Similar to the tensor predictor regression, Gaussian Process (GP) based nonparametric models are also studied for regression problems with tensor responses. Li et al.~\cite{Li2020} propose a method based on the Gaussian process regression networks (GPRN), where no special kernel structure is pre-assumed. Tensor/matrix-normal variational posteriors are introduced to improve the inference performance.

The aforementioned methods assume a low-dimensional structure of the predictors (either in the form of a  vector or a matrix), and are generally incapable of modeling high-dimensional tensor predictors. Under such circumstances, various tensor-on-tensor methods are proposed to deal with regression problems with both tensor-valued responses and predictors, and some are analyzed under the Bayesian framework. Given a tensor response $\mathcal{Y}_i\in\mathbb{R}^{p_1\times...\times p_K}$ and tensor predictors $\mathcal{X}_i\in\mathbb{R}^{m_1\times...\times m_K}$, Hoff~\cite{Hoff2015} associate $\mathcal{Y}_i$ and $\mathcal{X}_i$ through a Tucker structure (\ref{Tucker})
\begin{equation} 
\mathcal{Y}_i=\mathcal{X}_i\times_1\boldsymbol{B}_1\times_2\boldsymbol{B}_2\times_3\cdots\times_K\boldsymbol{B}_K+\mathcal{E}_i,
\label{eq21}
\end{equation}
where $\boldsymbol{B}_1,...,\boldsymbol{B}_K$ are matrices of dimension $p_1\times m_1,...,p_K\times m_K$ respectively. The error tensors $\mathcal{E}_i$ are i.i.d with dimension $p_1\times\cdots\times p_D$, and are assumed to follow a tensor normal distribution
$$ \mathcal{E}_i\sim \text{TN}(\boldsymbol{0};\boldsymbol{\Sigma}_1,...,\boldsymbol{\Sigma}_K).$$
Under the Bayesian framework, matrix normal priors are assigned to $\boldsymbol{B}_k|\boldsymbol{\Sigma}_k$, and inverse Wishart priors are imposed on $\boldsymbol{\Sigma}_k$ ($k=1,2,...,K$) to deliver efficient posterior computation.

Hoff~\cite{Hoff2015} require that the responses and predictors have the same number of modes. Lock~\cite{Lock} circumvent this restriction by employing a regression structure based on the tensor contraction product in \eqref{eq4}. Utilizing the same structure, Billio et al.~\cite{Billio2022} develop a Bayesian dynamic regression model that allows tensor-valued predictors and responses to be of arbitrary dimension.
Specifically, denote the tensor response by $\mathcal{Y}_t\in\mathbb{R}^{p_1\times...\times p_{D_1}}$ and the tensor predictor measured at time $t$ by $\mathcal{X}_t\in\mathbb{R}^{q_1\times...\times q_{D_2}}$. Billio et al.~\cite{Billio2022} propose the following dynamic regression model:
$$ \mathcal{Y}_t=\sum_{j=1}^q \mathcal{B}_j*\mathcal{Y}_{t-j}+\mathcal{A}*\mathcal{X}_t+\mathcal{E}_t,$$
where $\mathcal{B}_j$ and $\mathcal{A}$ are coefficient tensors of dimension $p_1\times\cdots\times p_{D_1}\times p_1\times\cdots\times p_{D_1}$ and $p_1\times\cdots\times p_{D_1}\times q_1\times\cdots\times q_{D_2}$, respectively, and $*$ is the tensor contraction product (\ref{contraction product}). The random error tensor $\mathcal{E}_t$ follows a tensor normal distribution, $ \mathcal{E}_t\sim\text{TN}(\boldsymbol{0}; \boldsymbol{\Sigma}_1,...,\boldsymbol{\Sigma}_{D_1}).$ The parsimony of coefficients is achieved by CP structures on the tensor coefficients, and an M-DGDP prior is assigned to the latent factors to promote shrinkage across tensor coefficients and improve computational scalability in high-dimensional settings.

\subsection{Theoretical Properties of Bayesian Tensor Regression}\label{Theo}
In this section, we discuss the theoretical properties for several  Bayesian tensor regression methods. 

In~\cite{Suzuki2015}, the in-sample predictive accuracy of an estimator coefficient tensor $\hat{\mathcal{W}}$ in (\ref{Suzuki}) is defined by
$$ \Vert \hat{\mathcal{W}}-\mathcal{W}^*\Vert_n^2:=\frac{1}{n}\sum_{i=1}^n \langle X_i,\hat{\mathcal{W}}-\mathcal{W}^*\rangle^2,$$
where $\mathcal{W}^*$ is the true coefficient tensor, $\{X_i\}_{i=1}^n$ are the observed input samples. Here $\Vert\cdot\Vert_n$ is not the usual $l_2$-norm. The out-of-sample predictive accuracy is defined by
$$ \Vert\hat{\mathcal{W}}-\mathcal{W}^*\Vert_{L_2(P(X))}^2:=E_{X\sim P(X)}[\langle X, \hat{\mathcal{W}}-\mathcal{W}^*\rangle^2],$$
where $P(X)$ is the distribution of $X$ that generates the observed samples $\{X_i\}_{i=1}^n$ and the expectation is taken with respect to $P(X)$. 

Assume that the $l_1$-norm of $X_i$ is bounded by $1$, the convergence rate of the expected in-sample predictive accuracy of the posterior mean estimator $\int \mathcal{W}d\Pi(\mathcal{W}|Y_{1:n})$,
$$ E\bigg[\bigg\Vert \int \mathcal{W}d\Pi(\mathcal{W}|Y_{1:n})-\mathcal{W}^*\bigg\Vert_n^2\bigg],$$
is characterized by the actual degree of freedom up to a log term. Specifically, let $d^*$ be the CP-rank of the true tensor $\mathcal{W}^*$, and $M_1,...,M_K$ be the dimensions for each order of $\mathcal{W}^*$, the rate is essentially
$$ O\left(\frac{\text{degree of freedom}}{n}\right)=O\left(\frac{d^*(M_1+\cdots+M_K)}{n}\right)$$
up to a log term and is optimal. Although the true rank $d^*$ is unknown, by placing a prior distribution on the rank, the Bayes estimator can appropriately estimate the rank and give an almost optimal rate depending on the true rank. In this sense, the Bayes estimator is adaptive to the true rank. Additionally, frequentist methods often assume a variant of strong convexity (e.g., a restricted eigenvalue condition~\cite{Bickel} and the restricted strong convexity~\cite{Negahban}) to derive a fast convergence rate of sparse estimators such as Lasso and the trace-norm regularization estimator. In contrast, the convergence rate in~\cite{Suzuki2015} does not require the strong-convexity assumption in the model.  

In terms of the out-of-sample predictive accuracy, the convergence rate achieved is also optimal up to a log term under the infinity norm thresholding assumption ($\Vert\mathcal{W}^*\Vert_\infty<R$, where $R>0$). Specifically, the rate is
$$ O\left(\frac{d^*(M_1+\cdots+M_K)}{n}(R^2\vee 1)\right)$$
up to a log factor.

Based on equation (\ref{Guhaniyogi1}), Guhaniyogi et al.~\cite{Guhaniyogi2017} prove the posterior consistency of the estimated coefficient tensor $\mathcal{B}$. Define a Kulback-Leibler (KL) neighborhood around the true tensor $\mathcal{B}_n^0$ as
$$\mathbb{B}_n=\bigg\{\mathcal{B}_n:\frac{1}{n}\sum_{i=1}^n \text{KL}\left(f(y_i|\mathcal{B}_n^0), f(y_i|\mathcal{B}_n)\right)<\epsilon\bigg\},$$
where $f(\cdot)$ is the glm density in \eqref{Guhaniyogi1}.
Let $\Pi_n$ be the posterior probability given $n$ observations, Guhnaiyogi et al.~\cite{Guhaniyogi2017} establish the posterior consistency by showing that
$$\Pi_n(\mathbb{B}_n^c)\to 0~~a.s.~\text{as }n\to\infty$$
under the probability measure induced by the $  \mathcal{B}_n^0$ when the prior $\pi_n(\mathcal{B}_n)$ satisfies a concentration condition. Based on this result, Guhaniyogi et al. further establish the posterior consistency for the M-DGDP prior in their study.

In a subsequent work~\cite{Guhaniyogi2017b}, the authors relax the key assumption in~\cite{Guhaniyogi2017} which requires that both the true and fitted tensor coefficients have the same rank in CP decomposition. Instead, the theoretical properties are obtained based on a more realistic assumption that the rank of the fitted tensor coefficient is merely greater than the rank of the true tensor coefficients. Under additional assumptions, the authors prove that the in-sample predictive accuracy is upper bounded by a quantity given below:
$$ E_{\mathcal{B}_n^0}\int\Vert \mathcal{B}_n-\mathcal{B}_n^0\Vert_n^2\Pi(\mathcal{B}_n|y_{1:n},X_{1:n})\leq AH_n/n,$$
where $H_n=o\{\log(n)^d\}$ and $A$ are positive constants depending on the other parameters. By applying Jensen's inequality
\begin{equation*}
E_{\mathcal{B}_n^0}[\Vert E(\mathcal{B}_n|Y_{1:n}, \mathcal{X}_{1:n})-\mathcal{B}_n^0\Vert_n^2]\leq E_{\mathcal{B}_n^0}\int\Vert \mathcal{B}_n-\mathcal{B}_n^0\Vert_n^2\Pi(\mathcal{B}_n|Y_{1:n}, X_{1:n}),
\end{equation*}
the posterior mean of the tensor coefficient, $E(\mathcal{B}_n|Y_{1:n}, X_{1:n})$, converges to the truth with a rate of order $n^{-1/2}$ up to a $\log(n)$ factor, which is near-optimal. Similar to Suzuki~\cite{Suzuki2015}, this result on convergence rate does not require a strong convexity assumption on the model. 

For the AMNR function defined in equation (\ref{AMNR}), Imaizumi and Hayashi~\cite{Imaizumi2016} establish an asymptotic property of the distance between the true function and its estimator. Let $f^*\in\mathcal{W}^{\beta}(\mathcal{X})$ ($\mathcal{W}^{\beta}(\mathcal{X})$ is the Sobolev space) be the true function and $\hat{f}_n$ be their estimator for $f^*$. Let $M^*$ be the rank of the true function. Then the behavior of the distance $\Vert f^*-\hat{f}_n\Vert$
strongly depends on $M^*$. Let $\Vert f\Vert_n$ be the empirical norm satisfying
$$\Vert f\Vert_n^2:=\frac{1}{n}\sum_{i=1}^n f(x_i)^2.$$
When $M^*$ is finite, under certain assumptions and for some finite constant $C>0$, by \cite{Imaizumi2016}, it follows that
$$ E\Vert\hat{f}_n-f^*\Vert_n^2\leq Cn^{-2\beta/(2\beta+\max_k I_k)},$$
where $\max_k I_k$ is the maximum dimension of the tensor predictor $\mathcal{X}$. This property indicates that the convergence rate of the estimator achieves the minimax optimal rate of estimating a function in $\mathcal{W}^{\beta}$ on a compact support in $\mathbb{R}^{I_k}$. The convergence rate of AMNR depends only on the largest dimension of $\mathcal{X}$.

When $M^*$ is infinite, by truncating $M^*$ at a finite value $M$, the convergence rate is nearly the same as the case of finite $M^*$, which is slightly worsened by a factor $\gamma/(1+\gamma)$~\cite{Imaizumi2016}:
$$ E\Vert\hat{f}_n-f^*\Vert_n^2\leq C(n^{-2\beta/(2\beta+\max_k I_k)})^{\gamma/(1+\gamma)}.$$

For the CATCH model in \eqref{CATCH1}-\eqref{CATCH3}, Pan et al.~\cite{Pan2018P} establish the asymptotic properties for a simplified model, where only the tensor predictor $\mathcal{X}$ is collected (the covariates $\boldsymbol{U}$ are not included). They define the classification error rate of the CATCH estimator and that of the Bayes rule as
\begin{gather*}
    R_n=\text{Pr}(\hat{Y}(\mathcal{X}^{\text{new}}|\hat{\mathcal{B}}_k,\hat{\pi}_k,\hat{\boldsymbol{\mu}}_k)\neq Y^{\text{new}}),\\
    R=\text{Pr}(\hat{Y}(\mathcal{X}^{\text{new}}|\mathcal{B}_k,\pi_k,\boldsymbol{\mu}_k)\neq Y^{\text{new}}), 
\end{gather*}
where $\hat{\mathcal{B}}_k, \hat{\pi}_k$ and $\hat{\boldsymbol{\mu}}_k$ are the estimated coefficients, and $\mathcal{B}_k, \pi_k$ and $\boldsymbol{\mu}_k$ are true coefficients. Under certain conditions, $ R_n\to R$
 with probability tending to 1. In other words, CATCH can asymptotically achieve the optimal classification accuracy. 

In~\cite{Yang2016}, Yang and Dunson establish the posterior contraction rate of their proposed classification model. Suppose that the data are obtained for $n$ observations $y^n=(y_1,...,y_n)^\top$ ($y_i\in\{1,2,...,d_0\}$), which are conditionally independent given $\boldsymbol{X}^n=(\boldsymbol{x}_1,...,\boldsymbol{x}_n)^\top$ with $\boldsymbol{x}_i=(x_{i1},...,x_{ip_n})^\top$, $x_{ij}\in\{1,...,d\}$ and $p_n\gg n$. Assume that the design points $\boldsymbol{x}_1,...,\boldsymbol{x}_n$ are independent observations from an unknown probability distribution $G_n$ on $\{1,2,...,d\}^{p_n}$. Denote
\begin{equation*}
d(P,P_0)=\int \sum_{y=1}^{d_0}|P(y|x_1,...,x_p)-P_0(y|x_1,...,x_p)|G_n(dx_1,...,dx_p),
\end{equation*}
where $P_0$ is the true distribution, and $P$ is the estimated distribution. Then under the given prior and other assumptions, it follows that
$$ \Pi_n\{P:d(P,P_0)\geq M\epsilon_n|y^n,\boldsymbol{X}^n\}\to 0~~a.s.,$$
where $\epsilon_n\to 0~(n\epsilon_n^2\to\infty, \sum_n\exp(-n\epsilon_n^2)<\infty)$, $M$ is a constant, and $\Pi_n(A|y^n,\boldsymbol{X}^n)$ is the posterior distribution of $A$ given the observations. Based on this result, Yang and Dunson~\cite{Yang2016} further prove that the posterior convergence of the model can be very close to $n^{-1/2}$ under some near low rankness conditions.

Among tensor response regression problems, Guha and Guhaniyogi~\cite{Guha2021} establish the convergence rate for predictive densities of their proposed SGTM model. Specifically, let $f^*(\mathcal{Y}|\boldsymbol{x})$ be the true conditional density of $\mathcal{Y}$ given $\boldsymbol{x}$ and $f(\mathcal{Y}|\boldsymbol{x})$ be the random predictive density for which a posterior is obtained.  Define an integrated Hellinger distance between $f^*$ and $f$ as 
$$ \mathcal{D}_H(f,f^*)=\sqrt{\int\int(\sqrt{f(\mathcal{Y}|\boldsymbol{x})}-\sqrt{f^*(\mathcal{Y}|\boldsymbol{x})})^2\nu_{\mathcal{Y}}(d\mathcal{Y})\nu_{\boldsymbol{x}}(d\boldsymbol{x})},$$
where $\nu_{\boldsymbol{x}}$ is the unknown probability measure for $\boldsymbol{x}$ and $\nu_{\mathcal{Y}}$ is the dominating measure for $f$ and $f^*$. For a sequence $\epsilon_n$ satisfying $0<\epsilon_n<1,\epsilon_n\to 0$, and $n\epsilon_n^2\to\infty$, under certain conditions it satisfies
$$ E_{f^*}\Pi_n\{\mathcal{D}_H(f,f^*)>4\epsilon_n|\{\mathcal{Y}_i,\boldsymbol{x}_i\}_{i=1}^n\}<4e^{-n\epsilon_n^2}$$
for all large $n$, where $\Pi_n$ is the posterior density. This result implies that the posterior probability outside a shrinking neighborhood around the true predictive density $f^*$ converges to $0$ as $n\to\infty$. Under further assumptions, the convergence rate $\epsilon_n$ can have an order close to the parametric optimal rate of $n^{-1/2}$ up to a $\log(n)$ factor.

\subsection{Posterior computation}\label{sec:64}
In terms of posterior inference methods, sampling methods such as MCMC and variational methods (e.g., Variational Expectation Maximization, Variational Inference, and Variational Bayes) are the two popular choices for Bayesian tensor analysis. MCMC is utilized in a majority of Bayesian tensor regression and some Bayesian tensor completion (decomposition) problems. The ergodic theory of MCMC guarantees that the sampled chain converges to the desired posterior distribution, and sometimes the MAP result is utilized to initialize the MCMC sampling for accelerating the convergence~\cite{Xiong2010, Gao2012}. In order to reduce the computational cost and adapt to different situations, batch MCMC and online MCMC are also used for posterior sampling~\cite{Hu2015, Hu2015b}. 

As an alternative strategy to approximate posterior densities for Bayesian models, variational inference  is very frequently employed in Bayesian tensor completion methods. These methods do not guarantee producing samples from the exact target density, but they are in general faster and more scalable to large datasets than MCMC are. In this category, Variational Expectation Maximization (VEM)~\cite{Xu2012, Zhe2015N, Zhe2016N, Zhe2018}, Variational Inference (VI)~\cite{Zhe2016b, Wang2020, Hu2022, Li2020}, and Variational Bayes (VB)~\cite{Schein2014, Hu2015, Zhao2015, Zhao2015T, Zhao2015b, Tillinghast2020, Long2020} are the classical choices, and the recently developed auto-encoding VB algorithm is employed to deal with intractable distributions~\cite{Liu2018, He2018}. Various studies have also adopted specific frameworks to reduce computational complexity (e.g., batch VB~\cite{Hu2015}, variational sparse Gaussian Processes~\cite{Tillinghast2020, Zhe2018, Zhe2016b, Wang2020}) and accommodate online or streaming data (e.g., online VB-EM~\cite{Zhe2015N}, streaming VB~\cite{Du2018, Zhang2018}, and Assumed Density Filtering/Expectation Propagation~\cite{Fang2021T, Fang2021, Pan2020, Fang2022}). Additionally, Bayesian tensor completion (regression) methods also utilize other methods including 
MLE~\cite{Pan2018P}, MAP~\cite{Zhao2014P} and EM~\cite{Rai2015, Hayashi2010}.

\section{Conclusion}\label{Conclusion}

In Bayesian tensor analysis, the unique data structure and its high dimensionality create  challenges in both computation and theory. Bayesian methods impose different decomposition structures on the tensor-valued data or coefficients to reduce the number of free parameters. While CP, Tucker and non-parametric decompositions are the most commonly used decomposition structures, other decompositions have received some attention under the Bayesian framework in recent years (e.g., tensor ring~\cite{Long2020}, tensor train~\cite{Hu2022}, neural~\cite{He2018}).  

A full Bayesian model requires the complete specification of a probabilistic model and priors over model parameters, both of which depends on the data type. For example, in tensor completion, when the tensor is continuous, the elements are usually assumed to follow a Gaussian distribution with the tensor mean following a decomposition structure~\cite{Xiong2010, Liu2018, Xu2012}. The Gaussian distribution can be extended to model the binary data through a link function~\cite{Gao2012}. In terms of count data, an element-wise Poisson distribution is often utilized to relate the decomposition structure to the tensor-valued data, and a Dirichlet or Gamma prior can be applied to latent factors or the core tensor to enforce the non-negativity in coefficients~\cite{Schein2014, Hu2015, Schein2016}.
For tensor regression problems, multivariate normal priors are placed over latent factors in the CP decomposition, with a Gaussian-Wishart prior on the hyper-parameters of the normal distribution to achieve conjugacy~\cite{Xiong2010, Chen2019b, Gao2012}. Specific priors on core tensor (e.g., the MGP prior~\cite{Rai2014, Rai2015}, the Gamma-Beta hierarchical prior~\cite{Hu2015}) or latent factors~\cite{Zhao2015T} in CP/Tucker structure can promote automatic rank inference by letting the posterior decide the optimal rank. Sparsity priors such as the M-DGDP prior~\cite{Guhaniyogi2017, Billio2022} and the M-SB prior~\cite{Guhaniyogi2021} are also popular choices for latent factors in the CP structure to promote low rankness, and local/global sparsity. Integrating robust, interpretable and computationally scalable Bayesian tensor methods with complex models (e.g., nonlinear machine learning, reinforcement learning, causal inference, and dynamic models) remains an interesting future direction.

Bayesian tensor regression has been widely used in applications, especially in medical imaging analysis (e.g., MRI and EGG), where high resolution spatially correlated data are produced. For both tensor-predictor and tensor-response regressions, there is a need to model tensor-valued coefficients, which is achieved by using CP/Tucker decomposition or nonparametric models that utilize Gaussian processes to model the non-linear relationship in the coefficient tensor. Posterior inference is conducted by Markov Chain Monte Carlo (MCMC) with Gibbs sampling, optimization based methods (e.g., variational Bayes), and streaming methods (e.g., expectation propagation). It is still of interest to develop scalable algorithms that accommodate  challenging settings such as streaming data analysis. 

In terms of theoretical studies, most of the existing work focus on (near-)optimal convergence rates for posterior distributions of the tensor coefficients in regression-related problems \cite{Suzuki2015, Guhaniyogi2017b, Imaizumi2016, Pan2018P, Yang2016, Guha2021}. There are still many open problems such as theoretical analysis for Bayesian tensor completion (and other tensor problems that we did not cover in this review) and convergence analysis of computational algorithms.

\bibliographystyle{plain}
\bibliography{ref.bib}

\end{document}